\documentclass[10pt,journal,compsoc]{IEEEtran}

\ifCLASSOPTIONcompsoc
  \usepackage[nocompress]{cite}
\else
  \usepackage{cite}
\fi

\ifCLASSINFOpdf  
  \usepackage[pdftex]{graphicx}
  \graphicspath{{./figures/}}
\else
  \usepackage[dvips]{graphicx}
  \graphicspath{{./figures/}}
  \DeclareGraphicsExtensions{.eps, .png, .pdf}
\fi
\ifCLASSOPTIONcompsoc
  \usepackage[nocompress]{cite}
\else
  \usepackage{cite}
\fi

\ifCLASSINFOpdf
  \usepackage[pdftex]{graphicx}
  \graphicspath{{./figures/}}
\else
  \usepackage[dvips]{graphicx}
  \graphicspath{{./figures/}}
  \DeclareGraphicsExtensions{.eps, .png, .pdf}
\fi

\usepackage{dblfloatfix}
\usepackage{url}

\usepackage{url,comment,cite}
\usepackage[cmex10]{amsmath}
\usepackage{amssymb,amsfonts,bbm}
\usepackage{array}
\usepackage{graphicx}
\usepackage{color}
\usepackage{bm}
\usepackage{cleveref} 
\usepackage{enumitem}
\usepackage{epstopdf}
\usepackage[table]{xcolor}
\usepackage{multirow}
\usepackage{hhline}
\usepackage{url}
\usepackage{mathabx}
\usepackage{booktabs}
\usepackage{tabularx}
\usepackage{caption}
\usepackage{booktabs}
\usepackage{siunitx}
\usepackage{arydshln}
\usepackage{textcomp}
\usepackage{comment}
\usepackage{soul}
\usepackage{color, colortbl}
\usepackage{flushend}

\definecolor{gray(x11gray)}{rgb}{0.75, 0.75, 0.75}

\def\ben{\begin{enumerate}}
\def\beq{\begin{equation}}
\def\beqa{\begin{eqnarray}}
\def\bit{\begin{itemize}}
\def\een{\end{enumerate}}
\def\eeq{\end{equation}}
\def\eeqa{\end{eqnarray}}
\def\eit{\end{itemize}}

\def\argmax{\mathop{\mathrm{arg~max}}\limits}

\newcommand{\ra}[1]{\renewcommand{\arraystretch}{#1}}

\newsavebox\dotbox
\sbox{\dotbox}{\(\displaystyle\bigodot\)}
\newlength{\dotheight}
\DeclareMathAlphabet{\mathpzc}{OT1}{pzc}{m}{it}
\DeclareFontFamily{OT1}{pzc}{}
\DeclareFontShape{OT1}{pzc}{m}{it}{<-> s * [1.200] pzcmi7t}{}
\DeclareMathAlphabet{\mathscr}{OT1}{pzc}{m}{it}

\makeatletter
\def\hlinewd#1{%
  \noalign{\ifnum0=`}\fi\hrule \@height #1 \futurelet
   \reserved@a\@xhline}
\makeatother

\begin{document}

\title{Estimation of Static and Dynamic Urban\\Populations with Mobile Network Metadata\vspace*{-7pt}}

\author{
	Ghazaleh Khodabandelou,
	Vincent Gauthier,
	Marco Fiore,~\IEEEmembership{Senior Member,~IEEE},
	Mounim El-Yacoubi   
	\thanks{
		G. Khodabandelou, V. Gauthier and M. El-Yacoubi are with SAMOVAR, Telecom SudParis, CNRS, Universit\'e Paris Saclay, 91000 \'Evry, France.
		E-mail: name.surname@telecom-sudparis.eu.\protect\\
		M. Fiore is with CNR, 10129 Torino, Italy.
		E-mail: marco.fiore@ieiit.cnr.it.\protect\\
	}
	\vspace*{-7pt}
}

\markboth{IEEE TRANSACTIONS ON MOBILE COMPUTING}%
{Khodabandelou \MakeLowercase{\textit{et al.}}: Estimation of Dynamic Urban Populations}

\IEEEtitleabstractindextext{%

\begin{abstract}
Communication-enabled devices routinely carried by individuals have become pervasive,
opening unprecedented opportunities for collecting digital metadata about the
mobility of large populations. In this paper, we propose a novel methodology for
the estimation of people density at metropolitan scales, using subscriber presence
metadata collected by a mobile operator. Our approach suits the estimation of static
population densities, \textit{i.e.}, of the distribution of dwelling units per urban
area contained in traditional censuses. More importantly, it enables the estimation
of dynamic population densities, \textit{i.e.}, the time-varying distributions of
people in a conurbation. By leveraging substantial real-world mobile network metadata
and ground-truth information, we demonstrate that the accuracy of our solution is
superior to that granted by state-of-the-art methods in practical heterogeneous
urban scenarios.
\vspace*{-4pt}
\end{abstract}


}

\maketitle

\IEEEdisplaynontitleabstractindextext

\section{Introduction}
\label{sec:INTRODUCTION}

Mobile network operators collect a profusion of metadata
from the digital communication activity of their subscribers. They are in a position to extract significant new knowledge on human behaviors at heterogeneous scales, ranging from single individuals to large populations. Examples abound, and are comprehensively reviewed in~\cite{naboulsi16survey}: they include original insights on mobility laws~\cite{song2010limits}, patterns of daily commuters~\cite{yang14}, dynamics of infective disease epidemics~\cite{bajardi2011human}, or people reaction to disaster situations~\cite{bagrow2011collective}.
Mobile network operators can then leverage such information to develop metadata-driven value-added services, for, \textit{e.g.}, transport analytics~\cite{telefonica} or location-based marketing~\cite{orange}.

Our work focuses on the use of mobile network metadata for the estimation of population density in urban regions, which is a paramount information for informed planning in metropolitan areas by local authorities. Traditional censuses are carried out at regular time intervals in all developed countries, and allow determining static population densities, \textit{i.e.}, the spatial distributions of citizens' places of habitation or \textit{dwelling units}. However, these censuses are complex to organize and expensive to run, which limits their periodicity to a few years in best cases~\cite{coleman13}. Instead, metadata collected by mobile network operators is fairly inexpensive to obtain, and covers large portions of the population~\cite{naboulsi16survey}. A reliable estimation of the static population density based on such metadata would eliminate the limitations of conventional survey-based approaches~\cite{ratti06,krings2009urban}.

In addition, mobile network metadata can be retrieved and analyzed with minimum latency. This paves the road to the characterization of dynamic population densities, \textit{i.e.}, of the instantaneous distributions of individuals that evolve as people move over time in the region of interest. The automated, near-real-time estimation of population density dynamics has inestimable value in supporting innovative services for instance to transport planning, large event organization, public safety, or law enforcement.

The utility of a reliable estimation of population distributions based on mobile network metadata has generated a flurry of recent studies on the subject~\cite{botta2015quantifying,kang2012towards,deville2014dynamic,douglass2015high,xu16}, which we review in Section~\ref{sec:RELATEDWORK}. Our work has the following advantages over such previous proposals.

\textit{(i)} Our design is based on a number of original metadata filters that significantly improve the accuracy of the population density estimate. The filters operate on subscriber presence, which we show to be a better proxy of the population density than other classes of mobile network metadata.

\textit{(ii)} We exploit a novel multivariate relationship of population density, subscriber presence and subscriber activity level for the estimation of dynamic population densities.

\textit{(iii)} When confronted with ground-truth data in multiple urban scenarios, our solution achieves good accuracy in estimating static and dynamic population densities. Specifically, it consistently outperforms current state-of-the-art approaches in both cases, and generates reasonable dynamic representations of the population distribution that allow, \textit{e.g.}, appraising attendance at sports and social events.

\textit{(iv)} Estimates of the dynamic population distribution obtained with our proposed model are openly accessible~\cite{zenodo}.
The datasets describe one month of population density fluctuations in the cities of Milan, Rome and Turin.

The paper is organized as follows. Section~\ref{sec:DATASET} describes our reference datasets. Section~\ref{sec:NIGHTTIME} presents the model for static population estimation, which is assessed in Section~\ref{sec:STATIC_EVALUATION}. Model refinements for the dynamic case are explained in Section~\ref{sec:DYNAMIC} and evaluated in Section~\ref{sub:DYNAMIC_EVALUATION}. Comparisons with the state-of-the-art are in Section~\ref{sec:COMPARATIVE}, and related works are reviewed in Section~\ref{sec:RELATEDWORK}. Section~\ref{sec:CONCLUSION} concludes the document.

\vspace*{-8pt}
\section{Datasets}
\label{sec:DATASET}

We leverage several datasets made available by Telecom Italia Mobile (TIM) within their 2015 Big Data Challenge~\cite{TIM2015}. We focus on three major conurbations in Italy, \textit{i.e.}, Milan, Turin and Rome. For each city, we retrieve metadata about the mobile traffic activity (presented in Section~\ref{sub:traffic}), as well as ground-truth census data on the local population distribution (Section~\ref{sub:pop}). We then infer information on land use from the mobile network metadata itself (Section~\ref{sub:landuse}).

\vspace*{-8pt}
\subsection{Mobile network metadata}
\label{sub:traffic}

The mobile network metadata provided by TIM covers the months of March and April 2015, and describes the volume of traffic divided by type (incoming and outgoing voice calls, incoming and outgoing text messages, and Internet sessions), as well as the approximate presence of subscribers.
These features are commonly available to mobile network operators from Call Detail Records compiled for billing purposes, hence they represent a sensible choice for developing an estimator of population density that is reusable.
All metrics are aggregated in time and space. In time, the metadata is totaled over 15-minute time intervals. In space, metrics are computed over an irregular grid tessellation, whose geographical cells do not overlap and have sizes ranging from 255$\times$325 m$^2$ to 2$\times$2.5 km$^2$. The number of cells is 1419 for Milan, 571 for Turin and 927 for Rome.

\begin{figure}[tb]
  \centering
  \includegraphics[width=0.9\columnwidth]{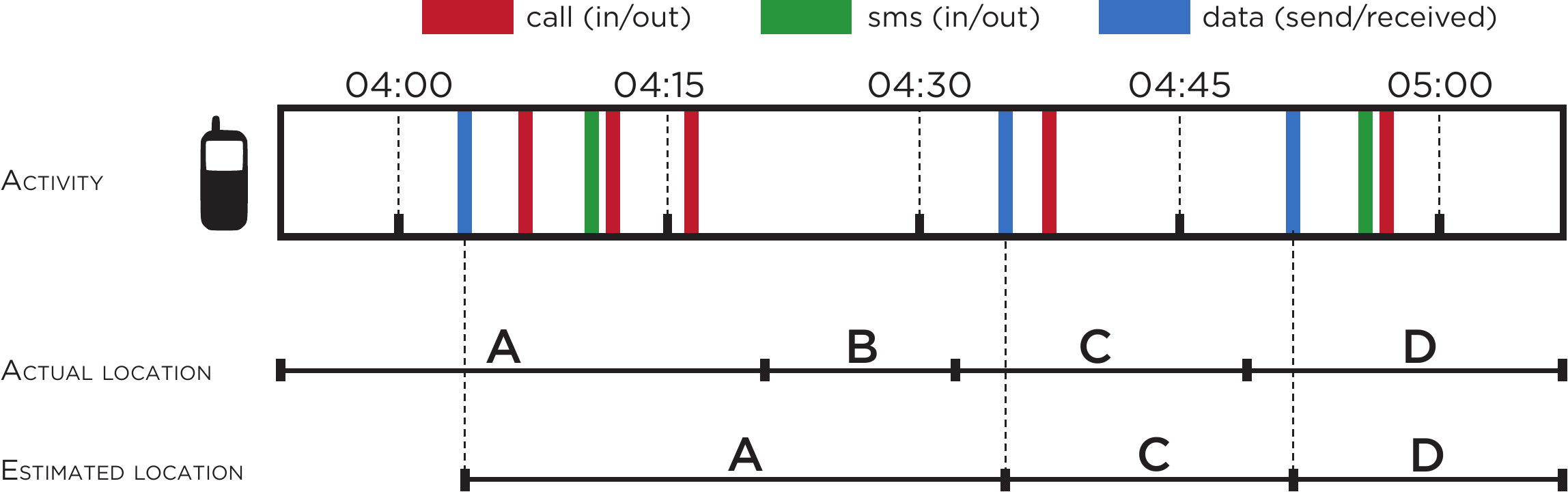}
  \caption{Example of subscriber location estimation.
	Network events of a mobile phone (top) allow approximating
	the cell where the user is located (bottom). This entails an
	error	with respect to the actual position of the user in time
	(middle).}
  \label{fig:presence}
  \vspace*{-7pt}
\end{figure}

While voice, text, and Internet traffic volumes are directly computed from the recorded demand, the presence information is the result of a simple preprocessing performed by the mobile network operator. Basically, each subscriber is associated to the geographical cell where he last interacted with the network for any purpose, which includes issuing or receiving a voice call, sending or receiving a text message, or establishing a new Internet data session. As each type of activity provides additional localization information, including them all in the calculation makes the presence information more accurate.
If a user is first detected within cell $A$, then he performs some activity in cell $B$ at time $t_1$, and his latest action is recorded in cell $C$ at time $t_2$, his location will be as follows: for $t \leq t_1$, the user is positioned in $A$; for $t_1 < t \leq t_2$, he is in $B$; for $t>t_2$, and until he performs an action in a different cell, he is in $C$. Figure~\ref{fig:presence} shows an example of this user location estimation process.
The presence metadata is then inferred by counting subscribers in each cell, at every 15 minutes.

\vspace*{-8pt}
\subsection{Population distribution}
\label{sub:pop}

Our ground-truth data for the static population distribution comes from the 2011 housing census run by ISTAT, the national organization for statistics in Italy.
It includes population counts, measured in terms of families, cohabitants, persons temporarily present, domiciles, and other types of lodging and buildings, for each administrative area.

To ensure spatial consistency between the mobile network metadata and the census data, we proceed as follows. Let us denote as $U_j$ the total number of inhabitants in the administrative area $j$, and as $A_j$ its surface. The population density $\rho_i$ in a geographical cell $i$, defined in Section~\ref{sub:traffic}, is then computed by assuming a uniform spatial distribution of inhabitants in each administrative area, as
\begin{equation}
\rho_i=\frac{1}{S_i}\sum_{j=1}^K U_j \frac{S_i \cap A_j}{A_j}
\label{eq:density}
\end{equation}
where $S_i$ is the surface of cell $i$, $K$ denotes the total number of administrative areas, and $S_i \cap A_j$ stands for the intersection surface of cell $i$ and administrative area $j$.

\vspace*{-8pt}
\subsection{Land use}
\label{sub:landuse}

Land use information is critical to the accurate estimation of population densities, and is regularly employed in the recent literature~\cite{douglass2015high,xu16}. We leverage the operator-collected data itself to classify the geographical cells based on their primary land use. To that end, we employ MWS, which is the current state-of-the-art technique for land use detection from mobile network metadata~\cite{furno16}.

MWS computes, for each spatial cell of the target region, a mobile traffic signature, \textit{i.e.}, a compact representation of the typical dynamics of mobile communications in the considered cell. Specifically, MWS signatures are computed as the median voice call and text activity in a cell recorded at every hour of the week. Signatures are clustered based on their shape, using a classical hierarchical algorithm on top of a correlation-based signature similarity measure. The output clusters group cells with similar types of human activities, \textit{i.e.}, belonging to a same land use. Averaging the signatures of all cells in a specific cluster allows defining \textit{characteristic signatures} associated to each land use, which are shown to be city-invariant: hence, once the characteristic signatures are identified, MWS can detect land use regions by solely using mobile network metadata.
When applied to our reference datasets, MWS classifies urban areas into five land uses: residential, office, touristic, university and shopping.

We stress that the land uses detected through mobile traffic metadata are more accurate and meaningful than those returned by other methods. As an example, the Milan conurbation territory is classified into buildings, vegetation, water, road, and railroads in~\cite{douglass2015high}: these categories, based on pure geographical features, have little relation with the activity of individuals. Instead, the classes identified by MWS in the same region, listed above, map to actual human endeavors, and, as such, have a stronger tie to the population density we aim at estimating.

\begin{figure}[tb] 
\centering
\vspace*{-6pt}
\hspace*{-2pt}
\begin{minipage}[b]{0.325\columnwidth}
\includegraphics[height=2.05\linewidth]{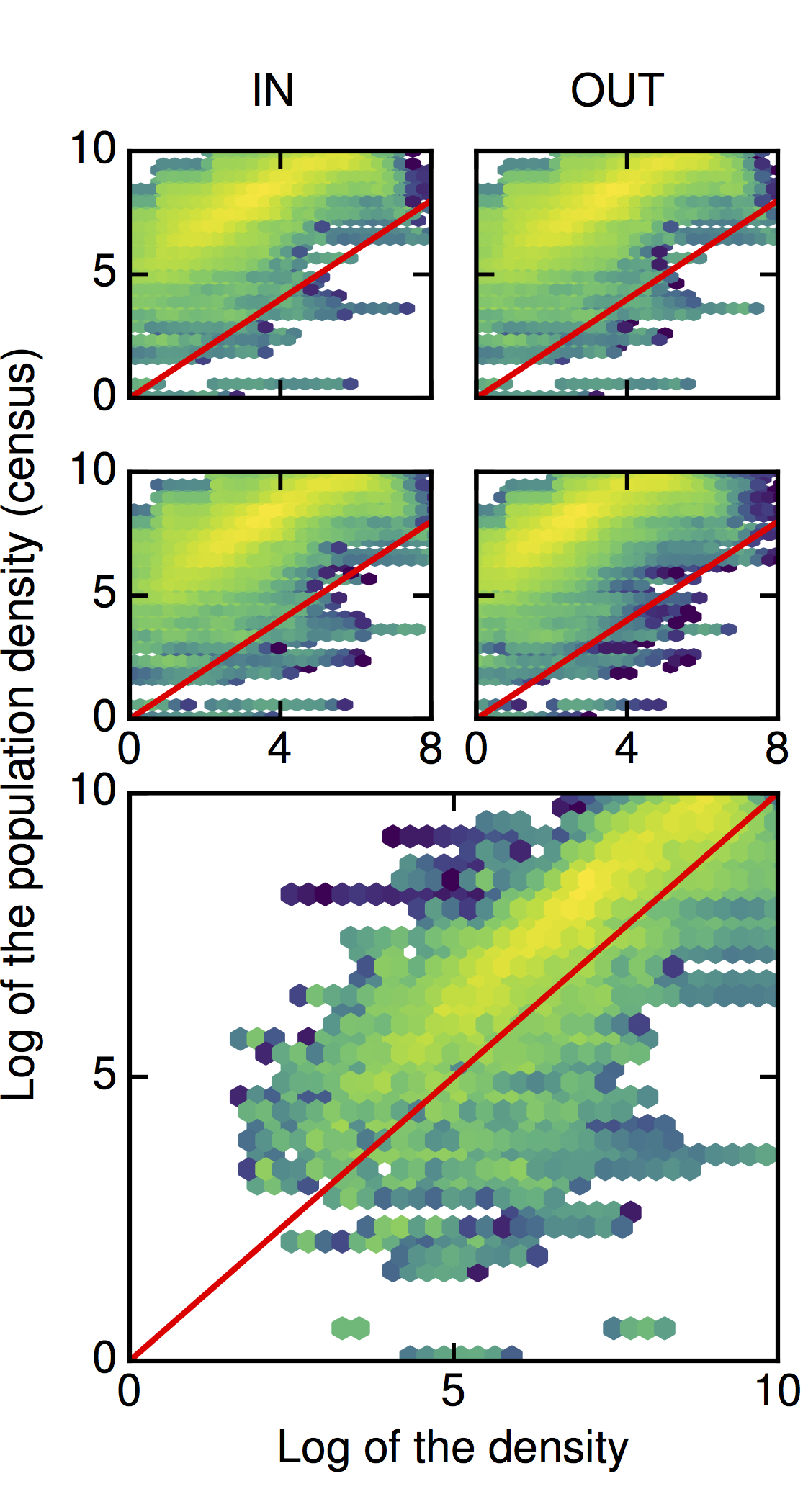}\vspace*{-3pt}
\centering{\small{Milan}}
\vspace{1ex}
\end{minipage}
\hspace*{0pt}
\begin{minipage}[b]{0.325\columnwidth}
\includegraphics[height=2.05\linewidth]{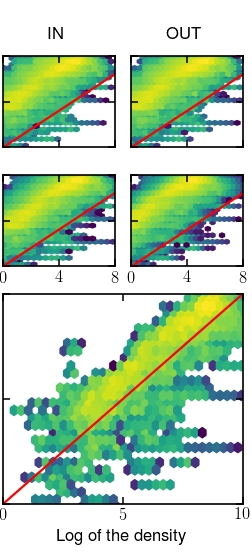}\vspace*{-3pt}
\centering{\small{Rome}}
\vspace{1ex}
\end{minipage} 
\hspace*{-6pt}
\begin{minipage}[b]{0.325\columnwidth}
\includegraphics[height=2.05\linewidth]{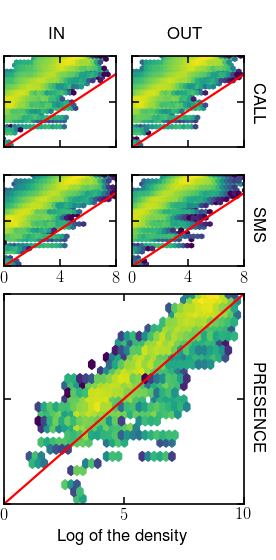}\vspace*{-3pt}
\centering{\small{Turin}}
\vspace{1ex}
\end{minipage}
\vspace*{-16pt}
\caption{ISTAT census population density versus density of calls, texts, and subscriber presence, in Milan, Rome, Turin.}
\label{fig:hetero}
\end{figure}

\vspace*{-8pt}
\section{Static population estimation}
\label{sec:NIGHTTIME}

Power laws regulate many relationships among social, spatial, and infrastructural properties of cities~\cite{bettencourt13}. In particular, previous works demonstrated the existence of a power relationship between the mobile network activity density $\sigma_i$ (\textit{i.e.}, activity per km\textsuperscript{2}) and the population density $\rho_i$ (\textit{i.e.}, inhabitants per km\textsuperscript{2}) in a region $i$~\cite{douglass2015high, deville2014dynamic}. In other words
\begin{equation}
\rho_i = \alpha \: \sigma_i^\beta.
\label{eq:regression}
\end{equation}
We find that such a power relationship holds in our reference scenarios as well.
Figure~\ref{fig:hetero} portrays heat-maps of the correspondence between the ground-truth population density and the network activity density, as recorded in each cell over the whole data collection period. The plots refer to different cities and different classes of mobile network metadata, \textit{i.e.}, incoming and outgoing calls and text messages, as well as presence.
All trends are clearly linear on a log-log scale: this implies that $log\left(\rho_i\right) = log(\alpha) + \beta log\left(\sigma_i\right)$, which is a simple transformation of (\ref{eq:regression}).

Our baseline population estimation model is thus described by the expression in (\ref{eq:regression}). By transforming the formula to a logarithmic scale as done above, we can use a linear regression model to estimate the parameters $\alpha$ and $\beta$.
Unfortunately, all linear relationships in Figure~\ref{fig:hetero} are characterized by a significant amount of noise, caused by the heterogeneity and heteroscedasticity of voice calls, text messages and subscriber presence density with respect to the actual population density.
Classical linear regression models assume absence of heteroscedasticity: running a regression directly on the raw metadata would yield poor results. In order to de-noise the metadata, we take a number of actions, which are discussed in the rest of the section.

\vspace*{-8pt}
\subsection{Metadata class filtering}
\label{sec:variablesfiltering}

\begin{table}
\centering 
\caption{Correlation coefficients between activity density of different types of mobile network metadata and the population census density, computed on aggregated data.\vspace*{-7pt}}
\ra{1.1}
\begin{tabular}{@{}llllllllll@{}}
& \multicolumn{2}{c}{Calls} & \multicolumn{2}{c}{Texts} & & \\
\cmidrule{2-3} \cmidrule{4-5}
& In & Out & In & Out & Internet & Presence \\
\midrule
\textsc{Milan} & 0.684 & 0.679 & 0.715 & 0.727 & 0.757 & \textbf{0.791} \\
\textsc{Rome} & 0.805 & 0.800 & 0.835 & 0.860 & 0.882 & \textbf{0.912} \\
\textsc{Turin} & 0.808 & 0.809 & 0.840 & 0.849 & 0.865 & \textbf{0.905} \\
\bottomrule 
\end{tabular}
\vspace*{-7pt}
\label{tab:corr}
\end{table}

As a first step, we determine which type of mobile network metadata is the most suitable to population estimation. To this end, we investigate the correlation between the population census data and the different classes of metadata presented in Section~\ref{sub:traffic}, as measured in each geographical cell over time and using the full two-month data.
Results are summarized in Table\,\ref{tab:corr}, which reports the Pearson correlation coefficient computed in the case of incoming and outgoing voice calls, incoming and outgoing text messages, Internet sessions, and presence. It is apparent that the presence is steadily better correlated than all other metadata, with coefficients of 0.79--0.91 that improve by 3.5\%--5\% the runner-up value associated with Internet session volumes.

\begin{figure*}[tb]
\centering
\begin{minipage}[b]{0.33\linewidth}
	\includegraphics[width=1.0\linewidth]{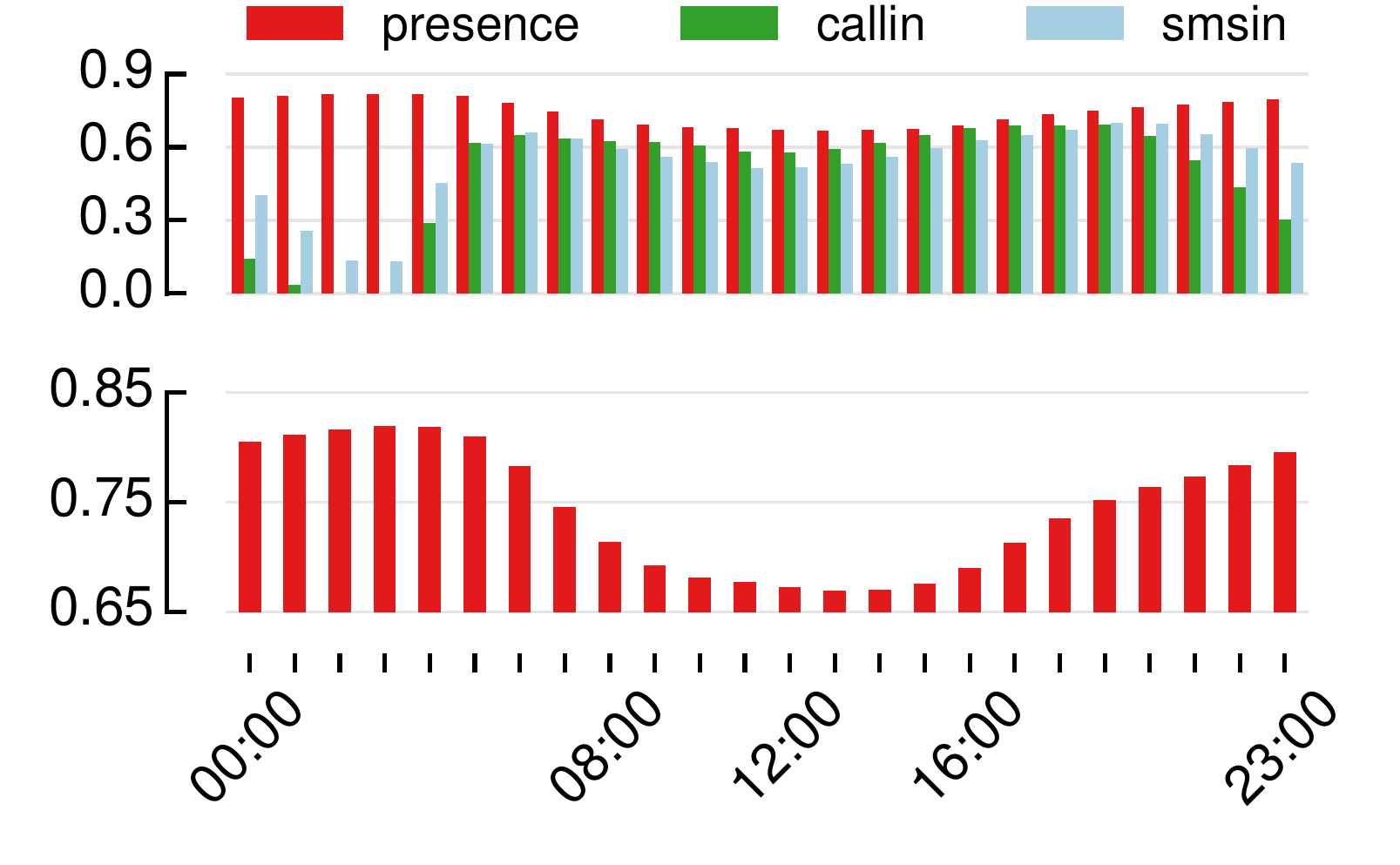}\vspace*{-6pt}
	\centering{\small{Milan}}
	\vspace*{-5pt}
\end{minipage}
\hspace*{-15pt}
\begin{minipage}[b]{0.33\linewidth}
	\includegraphics[width=1.0\linewidth]{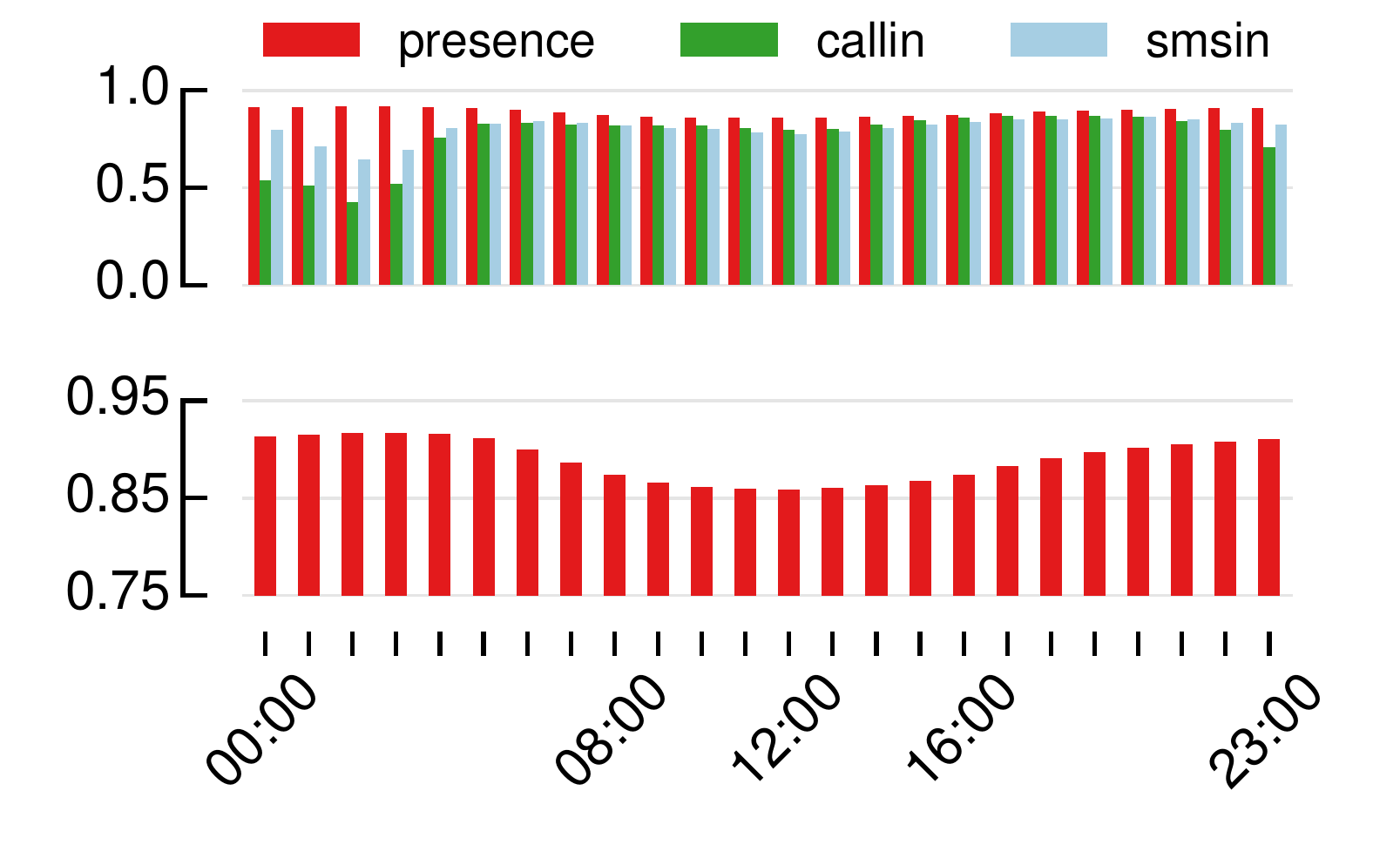}\vspace*{-6pt}
	\centering{\small{Rome}}
	\vspace*{-5pt}
\end{minipage} 
\hspace*{-15pt}
\begin{minipage}[b]{0.33\linewidth}
	\includegraphics[width=1.0\linewidth]{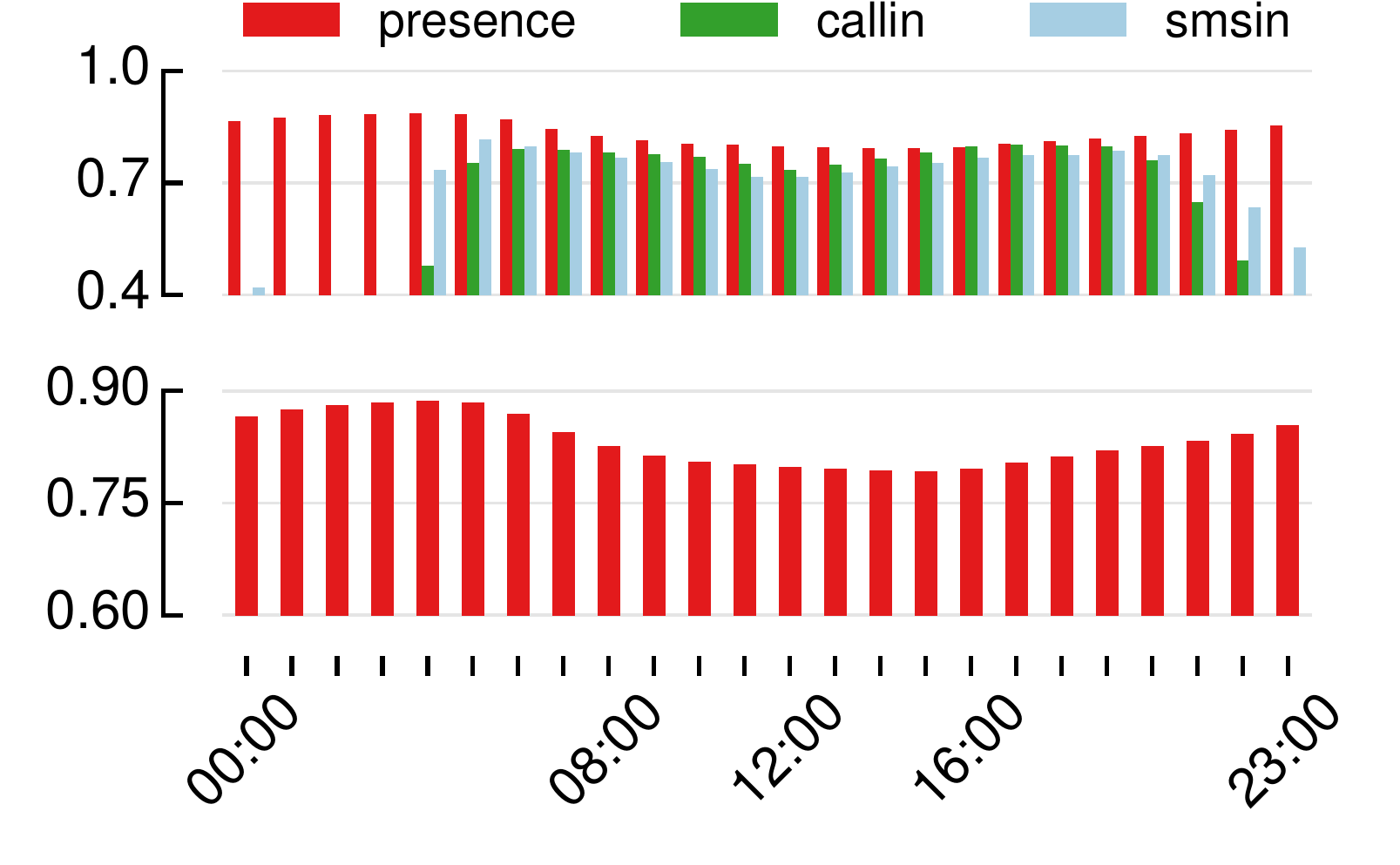}\vspace*{-6pt}
	\centering{\small{Turin}}
	\vspace*{-1pt}
\end{minipage}
\caption{Top: correlation between activity density of different types of mobile network metadata and the population census density, on a hourly basis. Bottom: zoom on presence metadata. Plots refer to Milan, Rome, Turin.}
\vspace*{-6pt}
\label{fig:type} 
\end{figure*}

\begin{figure*}[tb]
\centering
\begin{minipage}[b]{0.35\linewidth}
	\centering
	\includegraphics[width=0.95\linewidth]{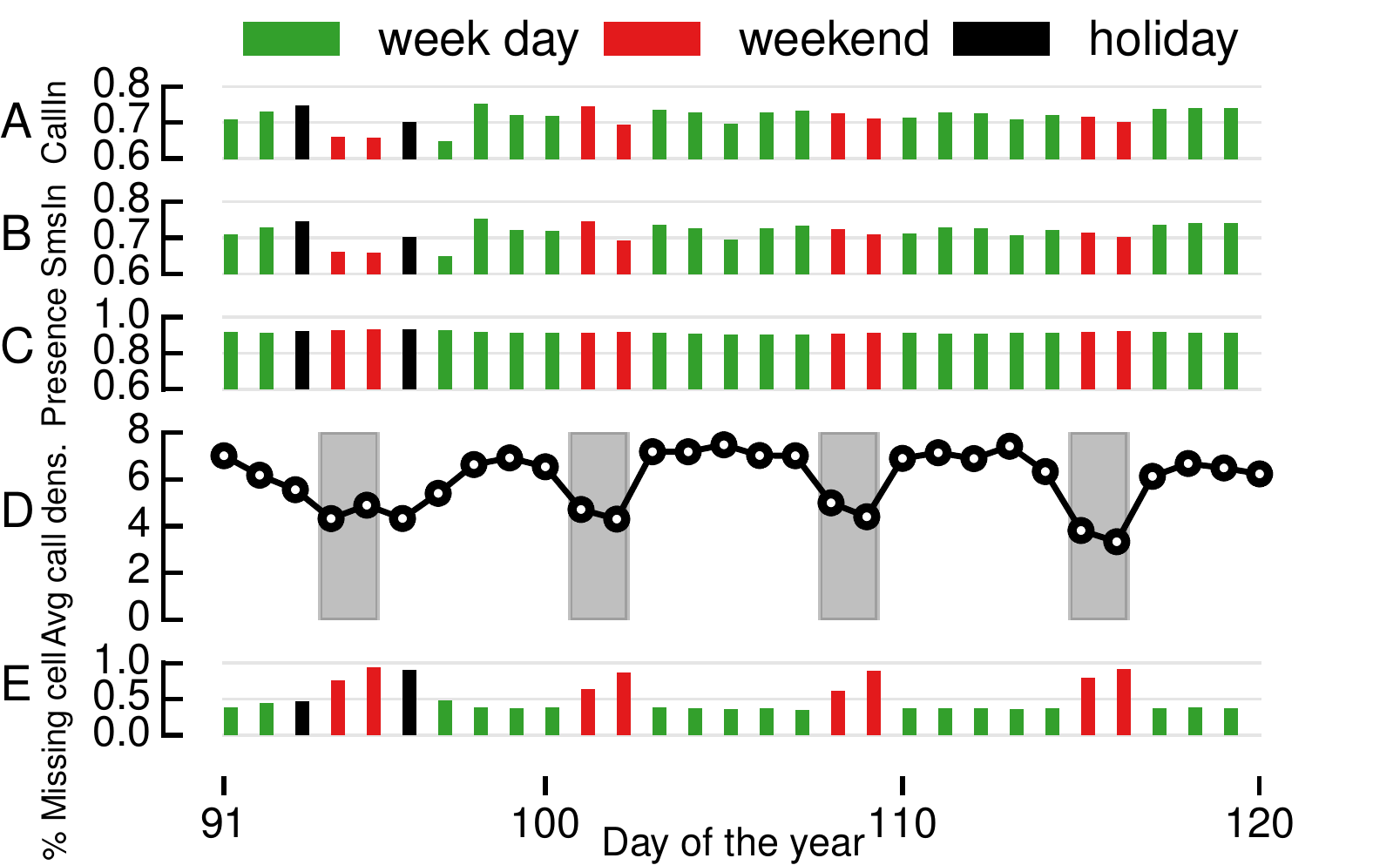}
	\centering{\small{Milan}}
	\vspace{1ex}
\end{minipage}
\hspace*{-26pt}
\begin{minipage}[b]{0.35\linewidth}
	\centering
	\includegraphics[width=0.91\linewidth]{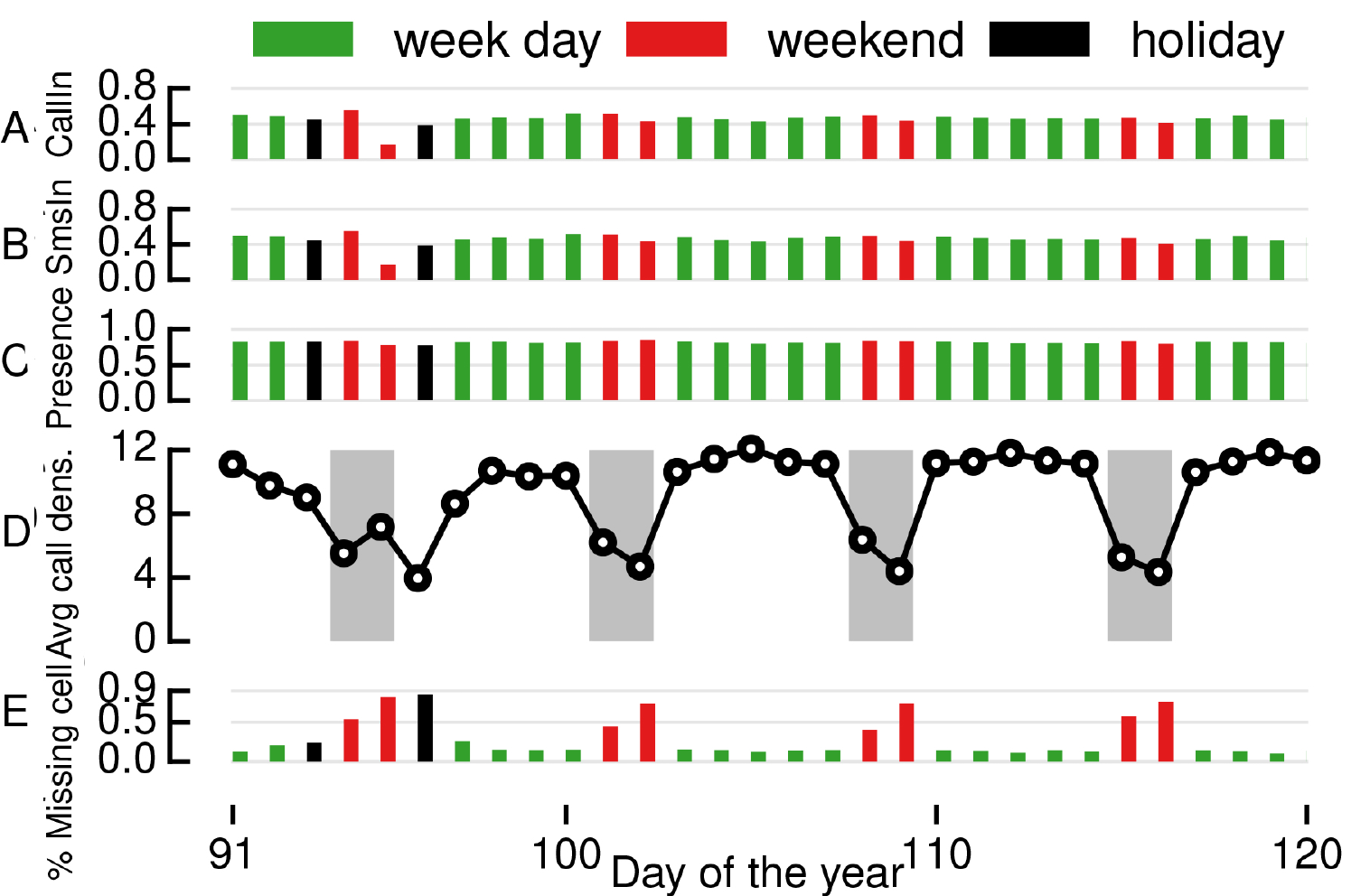}
	\centering{\small{Rome}}
	\vspace{1ex}
\end{minipage} 
\hspace*{-16pt}
\begin{minipage}[b]{0.35\linewidth}
	\centering
	\includegraphics[width=0.97\linewidth]{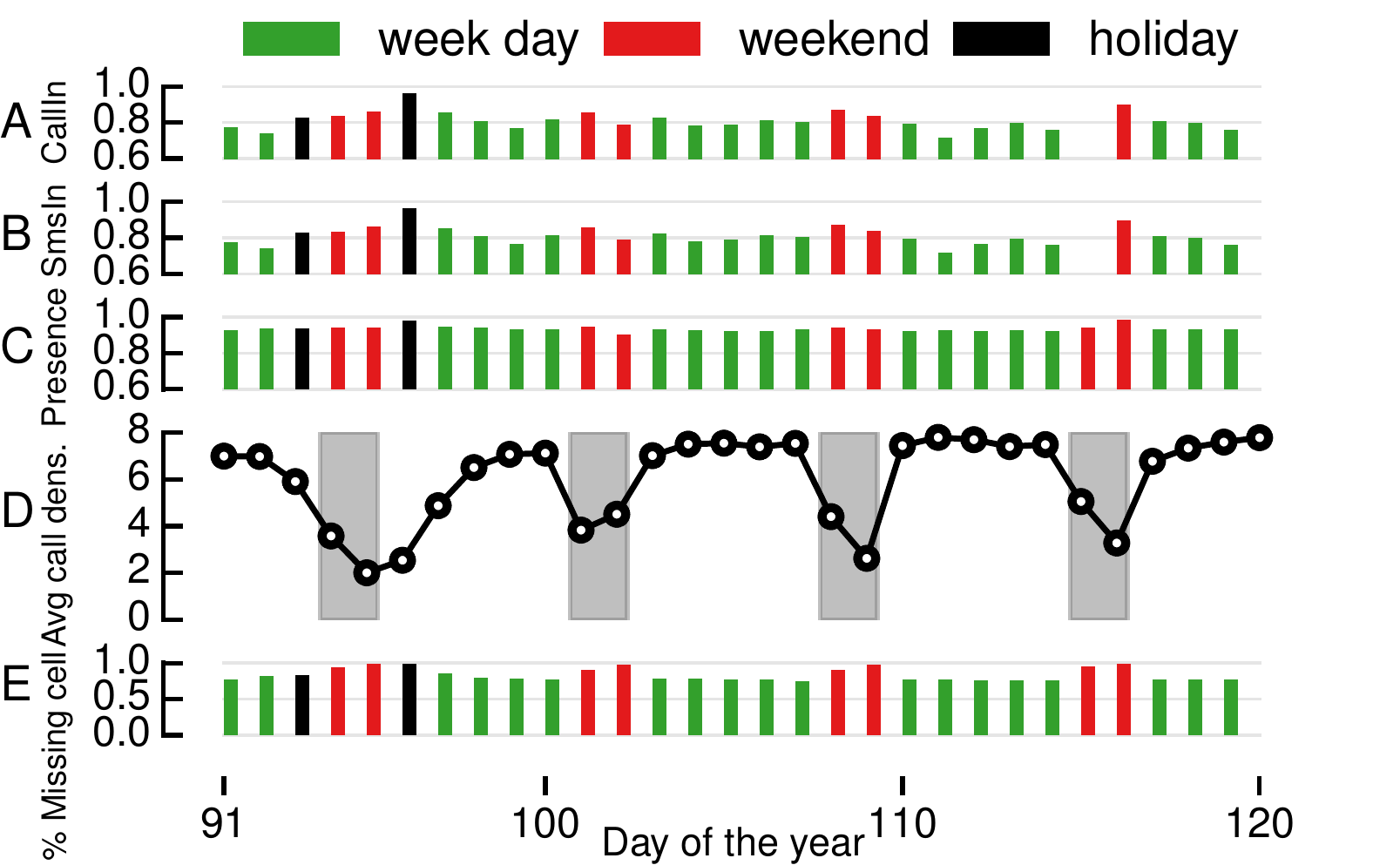}
	\centering{\small{Turin}}
	\vspace{1ex}
\end{minipage} 
\vspace*{-6pt}
\caption{Time dynamics of Milan, Rome and Turin metadata recorded between 4 am and 5 am in April 2015. A-C) Pearson correlation coefficient for incoming voice calls, incoming text messages, and subscriber presence. D) Average call density, with weekends highlighted. E) Percentage of cells with no subscriber presence metadata. Figure best viewed in colors.}
\label{fig:day}
\vspace*{-11pt}
\end{figure*}

We further examine the higher suitability of presence density as a proxy of population density in Figure~\ref{fig:type}. The top plots detail the typical daily fluctuation of the Pearson correlation coefficient, computed on 24 hourly aggregates of the two-month data. In addition to presence, the plots illustrate the correlation dynamics for incoming voice calls and text messages, as a benchmark%
\footnote{Outgoing calls and messages, and Internet sessions produced results equivalent to or worse than those obtained with incoming calls and messages. They are not shown here for the sake of clarity.}.
The results highlight that subscriber presence is sensibly better correlated with the ground-truth data, at all times and across all scenarios.

Interestingly, our results are aligned with those in~\cite{douglass2015high}, which, however, only considered calls and messages, and not the subscriber presence. In fact, their conclusion that calls between 10 am and 11 am yield the strongest correlation with population density, is superseded by our observation that subscriber presence is a much more relevant metric.
In the light of these considerations, we select the subscriber presence as the mobile network metadata on top of which we develop our methodology.

\vspace*{-8pt}
\subsection{Time filtering}
\label{sec:timefiltering}

A second dimension for filtering is time. As already observed in previous works, the correlation between mobile network metadata and population varies over time~\cite{deville2014dynamic}. This effect is also present in our case studies, as shown by the bottom plots in Figure~\ref{fig:type}: they detail the variability of the correlation coefficient for subscriber presence, over daytime and in the three reference urban scenarios.

The correlation coefficient has similar dynamics in all cities, and always peaks at night. The intuitive explanation is that the ISTAT census data refers to the static population density, and residents are more frequently at home overnight. This result allows selecting a single time filter that maximizes the correlation with the ground truth for all scenarios. Hereafter, and for the purpose of training our model, we will consider $\sigma_i$ in~(\ref{eq:regression}) as the subscriber presence density in cell $i$ during the interval between 4 am and 5 am.

\vspace*{-8pt}
\subsection{Day filtering}
\label{sec:daysfiltering}

As mentioned in Section~\ref{sub:traffic}, the mobile network metadata we employ covers 60 days in March and April 2015. An interesting question is if all these days should be considered in the regression model, or if there exists a more meaningful subset of days. Indeed, mobile network metadata is clearly affected by the diverse activity patterns and social phenomena that may characterize different days.

The three top plots in Figure~\ref{fig:day} show the Pearson correlation coefficient in the reference cities over April%
\footnote{Similar results for March are omitted here, for the sake of brevity.}.
They refer to incoming voice calls (A), incoming text messages (B), and subscriber presence (C); according to the previous discussion, they are computed over the 4 am to 5 am interval.
We still include calls and messages in order to check if they show correlation peaks on a daily basis that we could not observe in previous plots. This is not the case: the presence correlation coefficient ranges between 0.9 and 0.94, \textit{i.e.}, steadily higher values than those of calls and texts, lying between 0.6 and 0.8. We confirm that subscriber presence is a most convenient proxy of population distribution.
When it comes to filtering based on days, however, no clear trend is observed in the top three plots, for all metadata types.

A more insightful result in this sense is obtained by considering the percentage of cells for which no subscriber presence metadata is available between 4 am and 5 am on each day (E). Here, a remarkable weekly pattern emerges: high peaks of missing information appear on weekends (denoted by red bars) and holidays (denoted by black bars, and corresponding to Good Friday and Easter), and reach up to 90\% of cells. 
In order to understand why this happens, let us consider the average call density (D), \textit{i.e.}, the mean number of (incoming and outgoing) calls%
\footnote{Similar results were obtained with equivalent text message density and Internet session density, and are omitted for the sake of brevity.}
normalized by the cell surface and computed over all cells during each 15-minute slot between 4 am and 5 am: here, weekdays and weekends (highlighted in gray) yield remarkably different and lower density. Such a reduced communication activity leads to seldom updated presence metadata that easily misses user occupancy in less populated cells; therefore, it increases the number of low-presence cells, which are then removed from the dataset by the operator during sanitization to mitigate privacy risks%
\footnote{The rationale for the operator's choice is that if too few users are present in a cell, they may be tracked and re-identified in the metadata.}~\cite{TIM2015}. We conclude that a substantially lower activity of users during non-working days is the cause for the notable absence of presence metadata on those days.

In the light of these considerations, the high correlation between presence metadata and population census during weekends and holidays (C) is ostensible, as it only concerns cells for which metadata is available. For the (many) other cells, the correlation cannot be computed due to the lack of presence metadata. Since working with inconsistent temporal supports for different cells may introduce biases in the analysis, we ultimately opt for excluding weekends and holidays from the data used to derive our model.

\vspace*{-8pt}
\subsection{Outlier cell filtering}
\label{sec:RANSACRegression}

\begin{figure}[tb]
\centering
\begin{minipage}[b]{0.45\linewidth}
	\centering
	\includegraphics[width=1\linewidth]{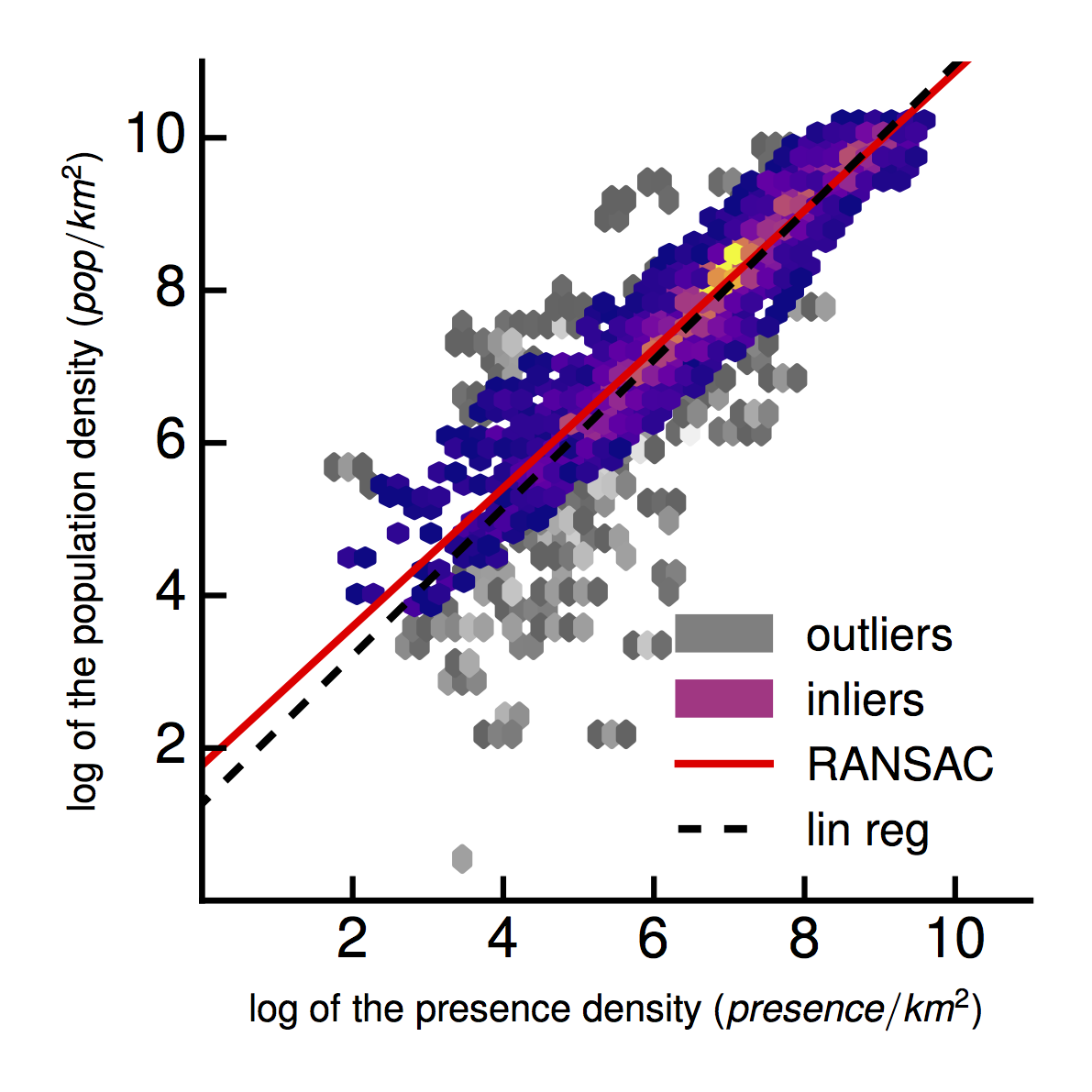}\vspace*{-3pt}
	\small{Milan}
\end{minipage}
\hspace*{-15pt}
\begin{minipage}[b]{0.45\linewidth}
	\centering
	\includegraphics[width=1\linewidth]{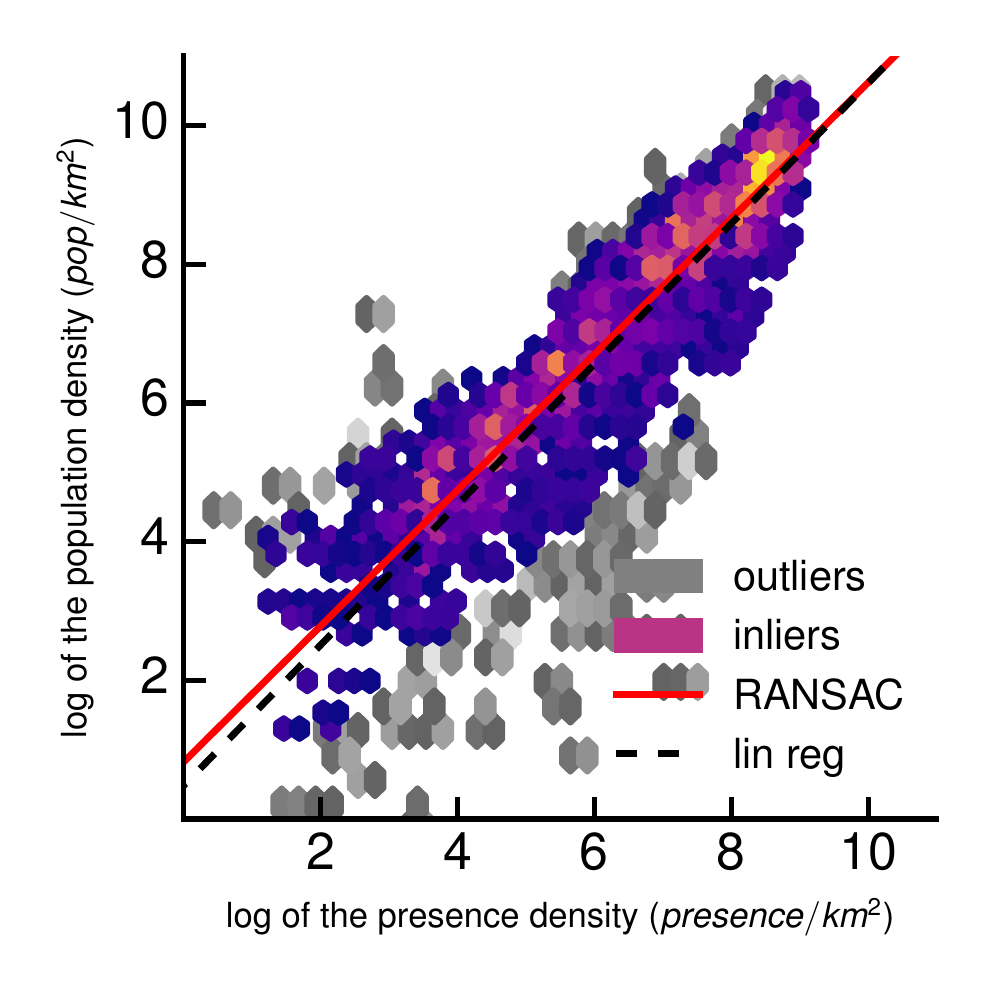}\vspace*{-3pt}
	\small{Rome}
\end{minipage}\\
\begin{minipage}[b]{0.45\linewidth}
	\centering
	\includegraphics[width=1\linewidth]{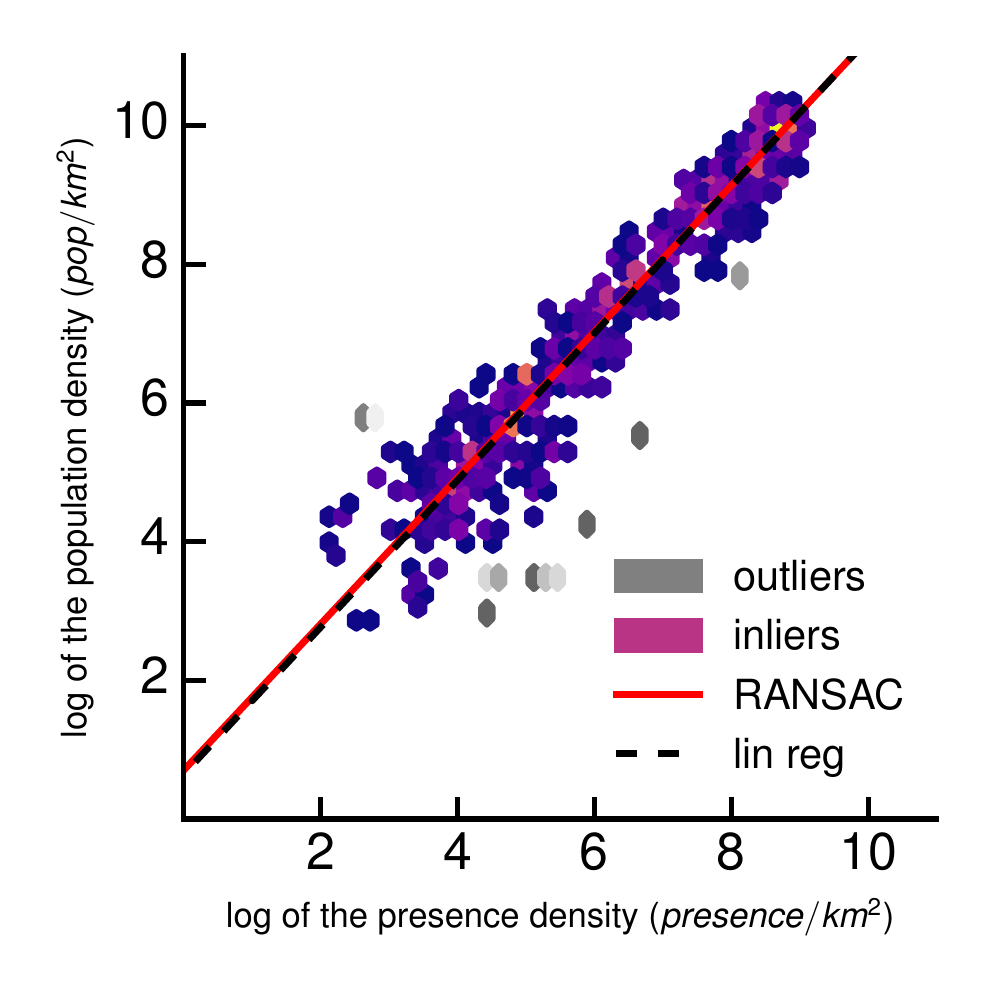}\vspace*{-3pt}
	\small{Turin}
\end{minipage} 
\hspace*{-15pt}
\begin{minipage}[b]{0.45\linewidth}
	\centering
	\includegraphics[width=1\linewidth]{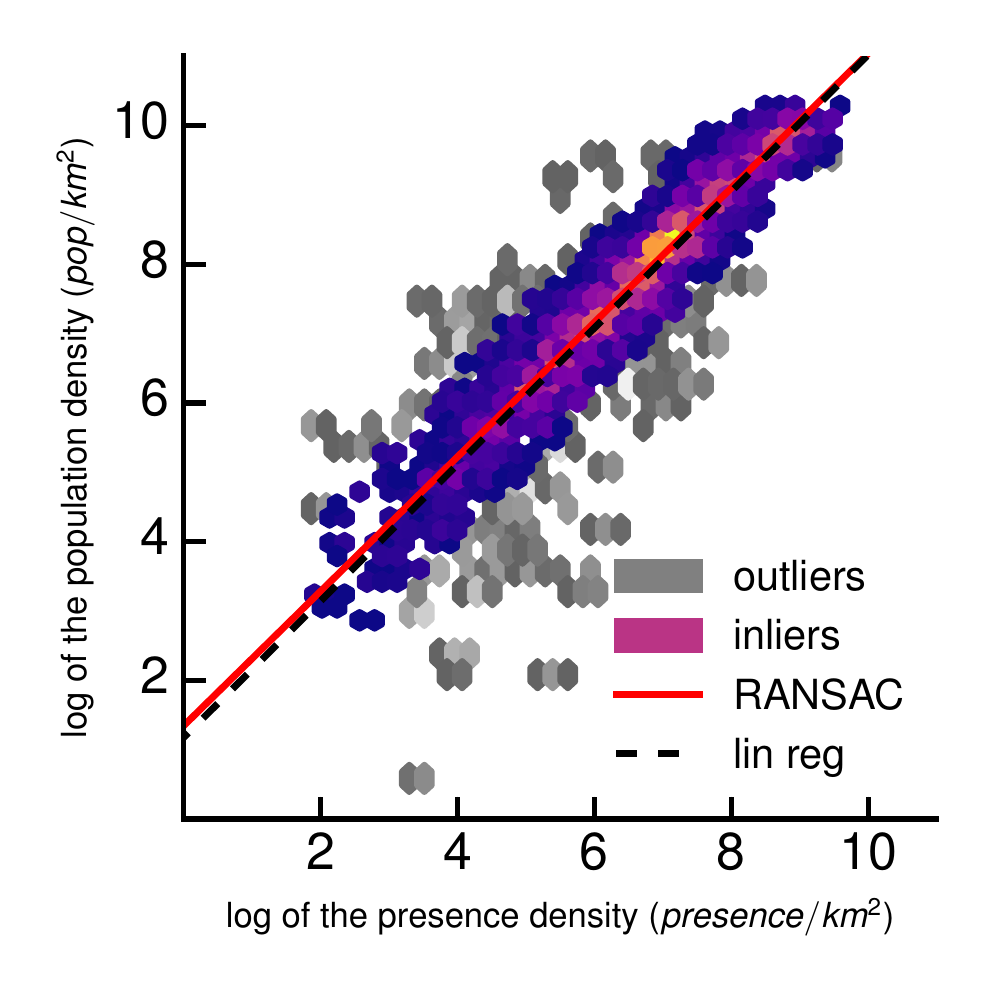}\vspace*{-3pt}
	\small{All cities}
\end{minipage} 
\caption{RANSAC regression on filtered subscriber presence metadata, in Milan, Rome, Turin, and combined scenarios.}
\label{fig:ransac1}
\vspace*{-2pt}
\end{figure}

In order to estimate the parameters $\alpha$ and $\beta$ in (\ref{eq:regression}), we employ the RANSAC regressor~\cite{fischler1981random} on the filtered subscriber presence metadata. In addition to estimating the model parameters in an iterative manner, RANSAC also automatically detects outlying points and excludes them from the regression.
Figure~\ref{fig:ransac1} depicts heatmaps of the filtered subscriber presence metadata versus the census population density data, for the three reference scenarios of Milan, Rome and Turin separately, as well as when the data for all these cities is considered at once. A first important observation is the effect of the proposed filters. The noise is sensibly reduced with respect to
the raw subscriber presence metadata, in Figure~\ref{fig:hetero}: most points are tagged as inliers (colored dots) by RANSAC, and follow a clear linear trend.

\begin{figure}[tb]
\centering
\hspace*{-4pt}
	\includegraphics[height=0.228\columnwidth]{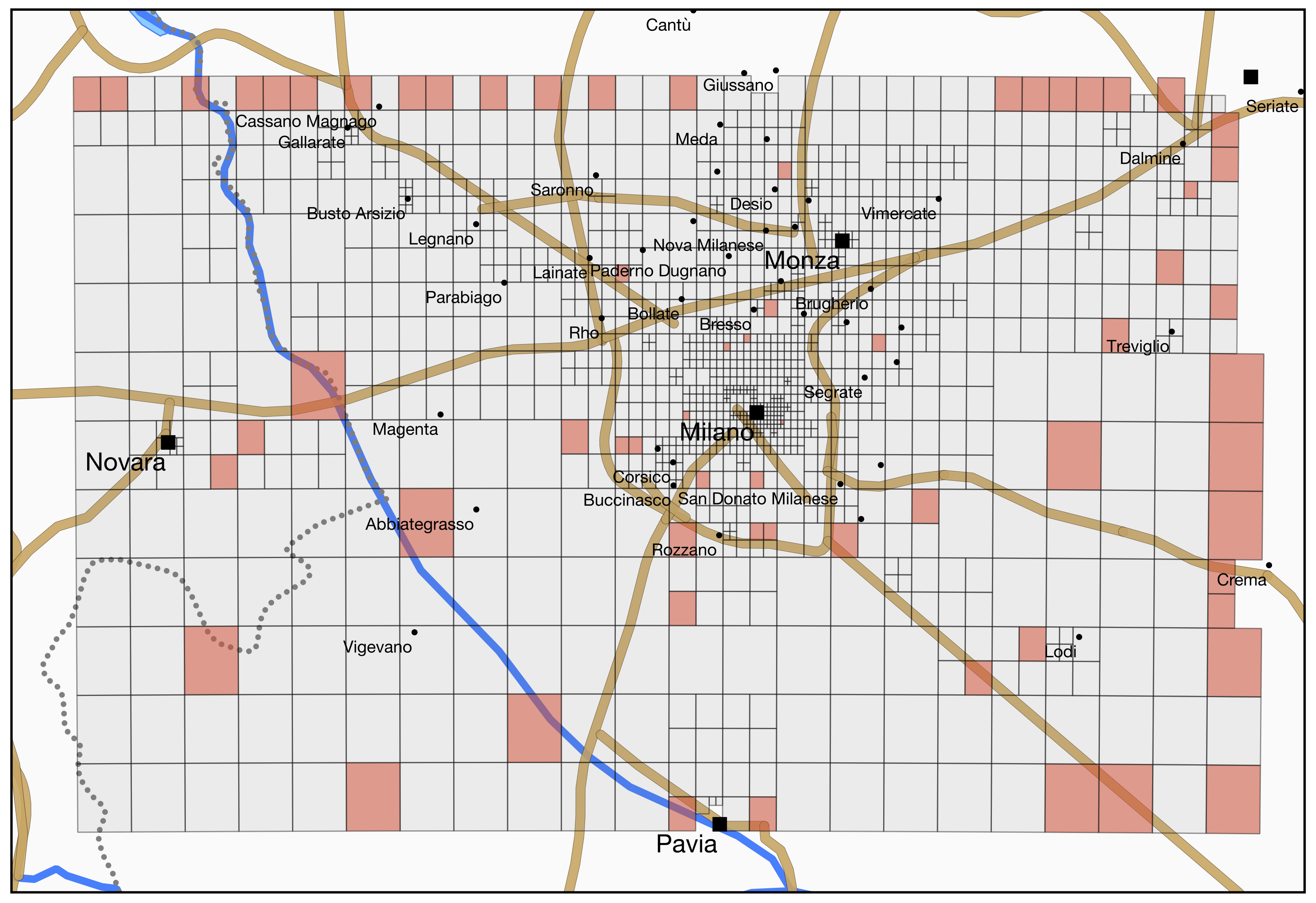}
\hspace*{-2pt}
	\includegraphics[height=0.228\columnwidth]{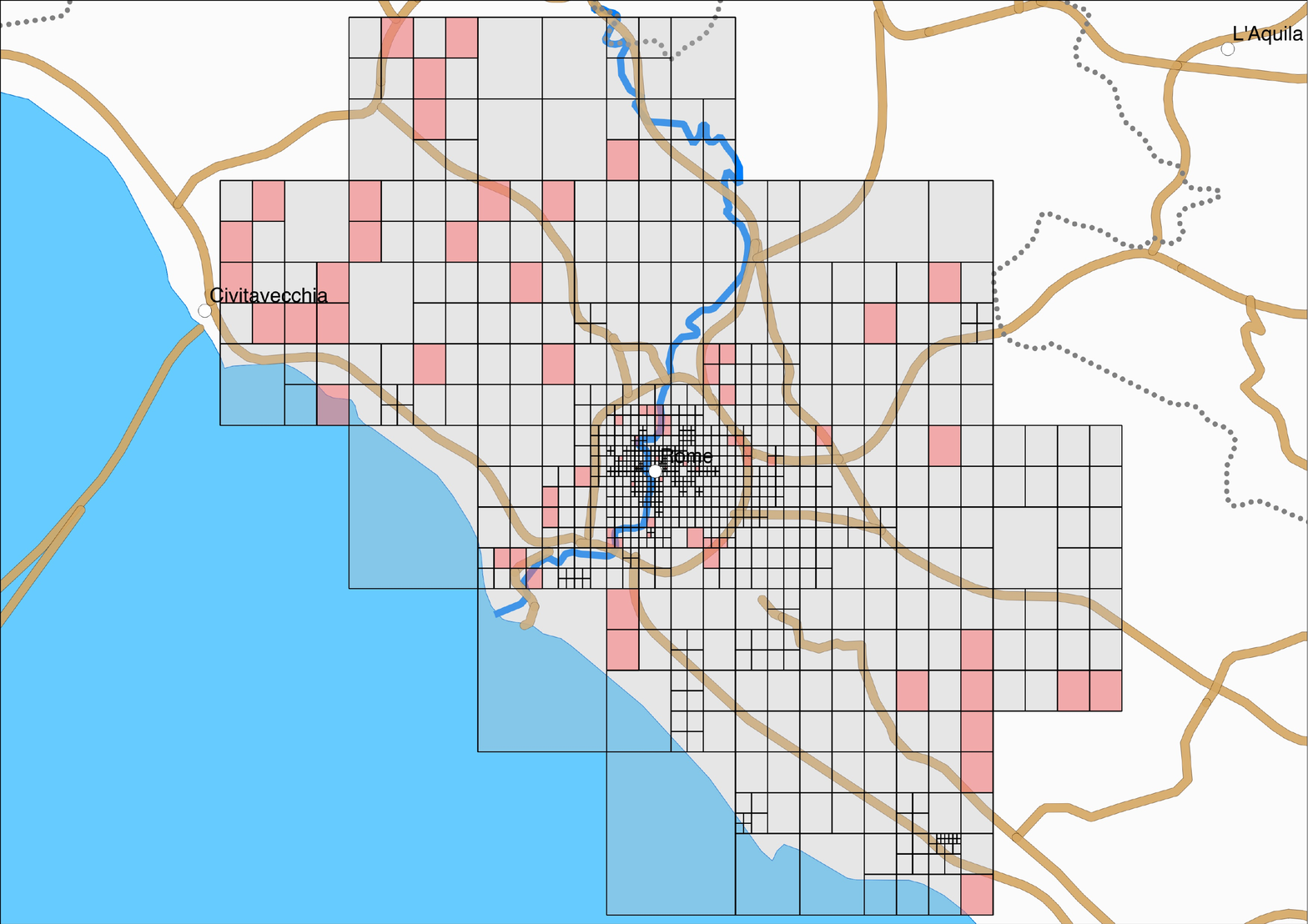}
\hspace*{-2pt}
	\includegraphics[height=0.228\columnwidth]{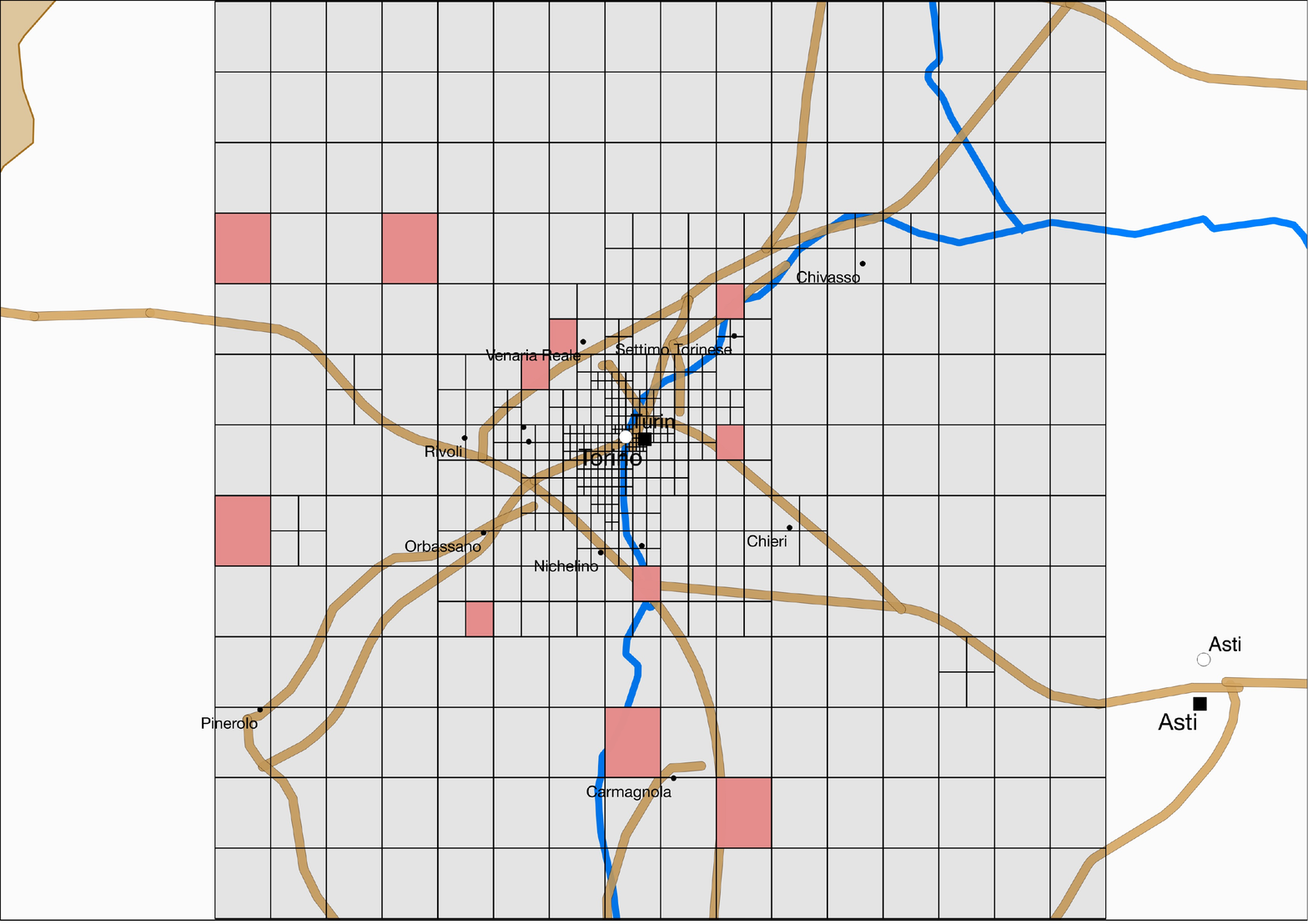}
\vspace*{-12pt}
\caption{Milan, Rome, Turin. Geographical positions of the cells that determine frequent outliers detected by RANSAC.}
\label{fig7:outliers1}
\vspace*{-12pt}
\end{figure}

Still, there exists a minority of outliers (gray dots) detected by RANSAC. A closer look at these outliers reveals that they refer to a same subset of cells, consistently over time. Thus, those cells yield some features that make the subscriber presence recorded there less related to the local amount of population.
Maps of such cells are in Figure~\ref{fig7:outliers1}.

We do not have strong evidence of why a limited number of specific cells show outlying behaviors with respect to the model. However, we speculate that two factors may contribute to this phenomenon. The first is a border effect: in all three cities, many outlying cells are located at the boundaries of the considered geographical region. We argue that the mobile network antennas in such cells probably cover areas (and populations) beyond the limits of the scenario, and the metadata reflects that. As a result, the subscriber presence associated to frontier cells also refers to users located outside the cells, which leads to an overestimation of the population by the model.

The second factor is the temporal mismatch between our datasets. The ISTAT census dates back to 2011, whereas the mobile network metadata refers to 2015. It is not unlikely that the distribution of the population density in several areas of the cities changed during the four years that separate the datasets. Figure~\ref{fig7:outliers1} seems to corroborate this hypothesis, since many outlying cells are located in suburban areas where resident population dynamics are more subject to evolve. If this were the case, mobile network metadata may thus help updating population distribution maps at a much higher frequency than traditional survey-based methods.

In summary, however, there is a risk that outlying cells may be affected by artifacts of the spatial tessellation or by potential issues in the associated ground-truth data; we thus deem safer to filter them out when training our model.

\vspace*{-8pt}
\section{Model evaluation}
\label{sec:STATIC_EVALUATION}

The regression returns fitted parameters $\hat{\alpha}$ and $\hat{\beta}$ of (\ref{eq:regression}). 
Our model estimates the static population density $\hat{\rho}_i$ in spatial cell $i$ from the filtered
subscriber presence density $\sigma_i$ as
\begin{equation}
\hat{\rho}_i =\hat{\alpha} \: \sigma_i^{\hat{\beta}}.
\label{eq:estimreg}
\end{equation}
Figure~\ref{fig:ransac1} illustrates the curves obtained with the model (solid red lines), as well as with a pure linear fitting where $\beta$=1 (dashed black lines). The two lines are close in all plots, which underscores the quasi-linearity of the relation between subscriber presence and population in all urban scenarios: \textit{e.g.,} we find $\hat{\beta}$=1.02 when all cities are used jointly.

\vspace*{-8pt}
\subsection{Metrics}
\label{sub:metrics}

In order to assess the accuracy of the model in (\ref{eq:estimreg}), we compute the determination coefficient ($R^2$), and the Normalized Root Mean Square Error ($NRMSE$) of the estimates, when compared to the ground-truth data.
The $R^2$ coefficient is
\begin{equation}\label{eq:r2}
R^2 = 1-\frac{\sum_{i=1}^N \left(\rho_i - \hat{\rho}_i\right)^2}{\sum_{i=1}^N \left(\rho_i-\bar{\rho}\right)^2},
\end{equation}
where $N$ denotes the number of cells in the spatial tessellation, and $\bar{\rho}$ is the average population density in all cells from ground-truth data.
We compute two versions of the $NRMSE$, which facilitates the comparison of the model results in different contexts.
They are defined as
\begin{equation}\label{eq:NRMSE1}
NRMSE(1) = \frac{1}{\rho_{max}-\rho_{min}}\;\sqrt{\frac{\sum_{i=1}^N(\hat{\rho}_i-\rho_i)^2}{N}},
\end{equation}
and
\begin{equation}\label{eq:NRMSE2}
NRMSE(2) = \frac{1}{\bar{\rho}}\;\sqrt{\frac{\sum_{i=1}^N(\hat{\rho}_i-\rho_i)^2}{N}},
\end{equation}
where $\rho_{max}$ and $\rho_{min}$ are the maximum and minimum population densities in cells within the target urban scenario, as indicated by the ground truth. In the following, we will use $NRMSE$ to refer to both expressions (\ref{eq:NRMSE1}) and (\ref{eq:NRMSE2}) at once.

We stress that our performance metrics are computed on all cells, including those excluded from model training.

\vspace*{-8pt}
\subsection{Milan case study}
\label{sub:milan}

We first focus on the Milan scenario. We adopt a three-fold cross-validation procedure, by separating the subscriber presence data into three equally-sized subsets in time. Two subsets are used as a training set to learn the model parameters, and the third is employed as a test set to evaluate the model quality. The process is repeated three times, by changing the test subset at each iteration.

\begin{figure}[tb]
  \centering
  \includegraphics[width=0.95\linewidth]{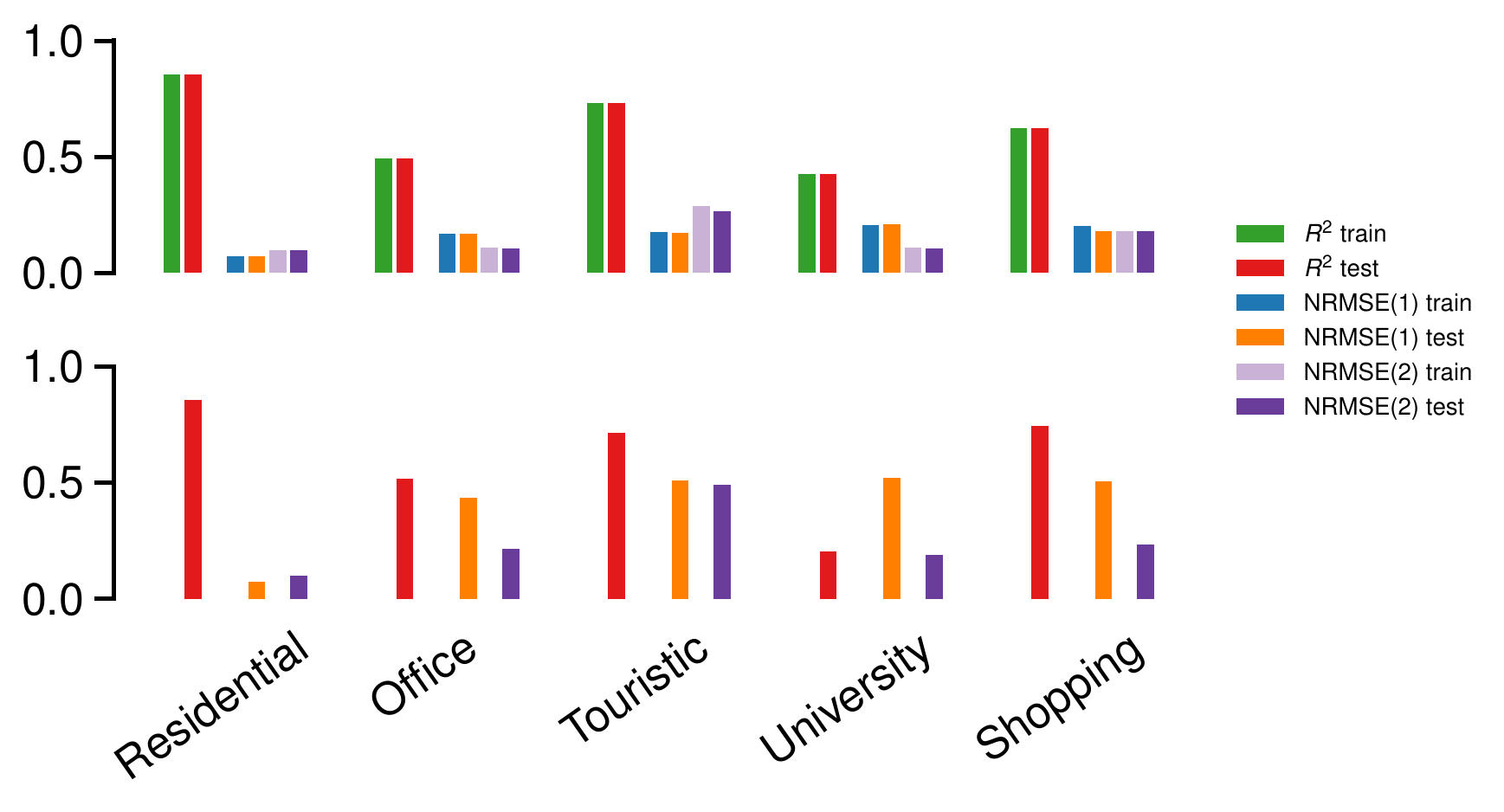}
	\vspace*{-4pt}
  \caption{Milan. Top: $R^2$ and $NRMSE$ for training and test sets, separated by land use. Bottom: $R^2$ and $NRMSE$ of the model trained on residential land use on other land uses.}
  \label{fig:corrclus} 
  \vspace*{-12pt}
\end{figure}

The baseline result for the Milan case study is shown in Figure~\ref{fig:corrclus}. The top plot reports the values of the $R^2$ and $NRMSE$ metrics computed on the test set, as well as on the training set for completeness. The results are separated per land use: \textit{i.e.}, the model is independently trained and tested on mobile network cells characterized by different land uses, extracted from the mobile network metadata as per Section~\ref{sub:landuse}. We remark that both $R^2$ and $NRMSE$: \textit{(i)} yield comparable values for all pairs of training and test data; \textit{(ii)} undergo significant variability across land uses.
The first result proves that the model can correctly estimate unknown population densities from subscriber presence metadata. The second highlights how land use affects the activities of individuals, including their mobile communication habits; in turn, this diversity influences the accuracy of population estimates from mobile network metadata.

The estimation is better in residential areas, where $R^2$=0.85, $NRMSE(1)$=0.075 and $NRMSE(2)$=0.122. This is a reasonable result, given that the distribution of dwelling units in the ISTAT census population is best captured in neighborhoods where households prevail. The quality of the estimation is still good ($R^2$ above 0.6) for touristic and shopping zones, fair ($R^2$ close to 0.5) for office areas, and bad ($R^2$ below 0.4) for university areas. Our speculation is that different percentages of individuals present in these areas overnight are not resident inhabitants in the census data (but, \textit{e.g.} tourists, students, or locals hanging out), hence do not appear in the population ground truth.

In the light of these considerations, the model parameters estimated in residential areas have the highest chance to be those actually linking population and presence metadata. We verify to what extent a model trained in residential zones can estimate populations in areas characterized by different land uses in the bottom plot of Figure~\ref{fig:corrclus}. In the Milan case study, this approach does not degrade performance significantly, with the sole exception of university areas that however represent a negligible minority of spatial cells.
This lets us consider a unifying model in the remainder of our study, by training (\ref{eq:estimreg}) on presence metadata from cells where residential land use is predominant.

\vspace*{-8pt}
\subsection{Generalization to different city scenarios}
\label{sub:all-cities}

\begin{table*}[tb]
\renewcommand{\arraystretch}{1.1}
\setlength{\tabcolsep}{2pt}
\caption{Model parameters and accuracy in Milan, Rome, and Turin.}
\vspace*{-5pt}
\centering
\begin{tabular}{@{}llllllllllllllllll@{}}
& \multicolumn{12}{c}{\textsc{Residential}} & \phantom{abc} & \multicolumn{3}{c}{\textsc{Mixed}} \\
\cmidrule{2-13} \cmidrule{15-17}
&& \multicolumn{6}{c}{Training}  && \multicolumn{4}{c}{Test} &&& \\
\cmidrule{2-9} \cmidrule{11-13}
& $\hat{\alpha}$ & $95\%\text{ C.I.}$ & $\hat{\beta}$ & $95\%\text{ C.I.}$& &$R^2$ & $NRMSE(1)$ & $NRMSE(2)$ & &$R^2$ & $NRMSE(1)$ & $NRMSE(2)$ && $R^2$ & $NRMSE(1)$ & $NRMSE(2)$ \\
\midrule
\textsc{Milan} &3.45&[2.80,3.93]&0.97 &[0.89,1.0]&&0.86 & 0.073 & 0.097 && 0.85 & 0.075 & 0.097 && 0.80 & 0.087 & 0.122 \\
\textsc{Rome} &2.55&[2.27,3.32]&0.99&[0.88,1.2]&& 0.80 & 0.100 & 0.094 && 0.80 & 0.102 & 0.095 && 0.73 & 0.093 & 0.132  \\
\textsc{Turin} &2.11&[1.95,2.66]&1.03&[0.92,1.1]&& 0.87 & 0.088 & 0.048 && 0.87 & 0.081 & 0.039 && 0.84 & 0.087 & 0.187  \\
\bottomrule  
\label{tab:cities}
\vspace*{-17pt}
\end{tabular}
\end{table*}

Table\,\ref{tab:cities} summarizes the performance of our model in all urban scenarios. The {\it residential} portion of Table\,\ref{tab:cities} refers to results obtained with data relative to cells with residential land use. It shows: \textit{(i)} the model parameter values $\hat{\alpha}$ and $\hat{\beta}$ returned by fittings on the training set, with 95\% confidence intervals; \textit{(ii)} the fitting quality of the model computed over the training set, as $R^2$ and $NRMSE$; \textit{(iii)} the accuracy of the estimation in the test set, as $R^2$ and $NRMSE$.

We remark that the values of $\hat{\beta}$ are consistently close to one across all of the urban scenarios we consider. Instead, $\hat{\alpha}$ tends to be city-specific%
\footnote{We explored if a number of features (\textit{e.g.}, total population, average population density, conurbation size, number of cells in the geographical surface, mobile network operator market share, etc.) could explain the difference, without finding significant correlations.}.
The accuracy of the estimation is in all cases very good, attaining determination coefficients between 0.8 and 0.87, and a normalized error below 0.1.

The right portion of Table\,\ref{tab:cities}, under the {\it mixed} tag, outlines the performance of the model trained on residential areas, and then used to estimate the population density in the complete urban region, including zones that are not residential in nature.
The accuracy remains good%
\footnote{We remark that these values are aligned with or better than those of current state-of-the-art solutions for static population density estimation. For instance, $R^2$ of 0.66 and 0.8 reported in~\cite{douglass2015high} and in~\cite{xu16}, respectively; also, the $NRMSE$ is measured at 1.0 in~\cite{xu16}. A complete comparative evaluation is provided in Section~\ref{sec:COMPARATIVE}.},
with $R^2$ in the range between 0.76 and 0.82, and $NRMSE$ around 0.1.

\begin{table}
\renewcommand{\arraystretch}{1.1}
\setlength{\tabcolsep}{1.95pt}
\centering 
\caption{Models trained on data of cities on rows are used to estimate the population density of cities on columns.}
\begin{tabular}{@{}lllllllllllll@{}}
&& \multicolumn{11}{c}{\textsc{Mixed Land Use}} \\
\cmidrule{3-13}
&& \multicolumn{3}{c}{$R^2$} && \multicolumn{3}{c}{$NRMSE(1)$} && \multicolumn{3}{c}{$NRMSE(2)$} \\
\cmidrule{3-5} \cmidrule{7-9} \cmidrule{11-13}
&& Milan & Rome & Turin && Milan & Rome & Turin && Milan & Rome & Turin \\
\midrule
\textsc{Milan} && 0.80 & 0.67 & 0.84 && 0.087 & 0.102 & 0.088 && 0.122 & 0.143 & 0.186 \\
\textsc{Rome} && 0.82 & 0.73 & 0.78 && 0.083 & 0.093 & 0.103 && 0.121 & 0.132 & 0.179 \\
\textsc{Turin} && 0.77 & 0.64 & 0.84 && 0.094 & 0.108 & 0.087 && 0.124 & 0.139 & 0.187 \\
\bottomrule 
\end{tabular}
\vspace*{-12pt}
\label{tab:cross_city} 
\end{table}

As a final test, we explore the possibility of estimating the population density in an urban area using a model trained on data collected in another city. The rationale is that, in Table\,\ref{tab:cities}, the $\hat{\beta}$ values are almost identical for all cities, and $\hat{\alpha}$ values are not dramatically different.
Table\,\ref{tab:cross_city} summarizes the estimation accuracy on all possible combinations of our three reference city scenarios, for both $R^2$ and $NRMSE$. The $(i,j)$-th element in the table maps to the accuracy of a model trained on metadata and ground truth from city $i$, and used to estimate the population in city $j$.

The metric values stay fairly high (in the range 0.64-0.84) for $R^2$ and low (between 0.1 and 0.2) for $NRMSE$, in all combinations of cities, which suggests that a cross-city estimation of population density is in fact possible. This result has important practical implications, since it paves the road to the estimation of populations in cities using mobile network metadata, and without any need for training on ground-truth data on a specific urban region.

As a concluding remark, we underscore that all results above hold for our three reference cities, and we cannot claim generality beyond these. Yet, the strong consistence of model performance and the possibility of cross-city estimation are especially encouraging when considering that the target cities have fairly diverse topological features and population sizes, at approximately 2,850,000 (Milan), 1,350,000 (Rome) and 800,000 (Turin) inhabitants.

\vspace*{-8pt}
\section{Dynamic population estimation}
\label{sec:DYNAMIC}

An interesting possibility offered by mobile network metadata is the estimation of dynamic populations, \textit{i.e.}, the instantaneous distribution of city inhabitants over time, as determined by their daily activities. In this case, we do not target the evaluation of the static density of dwelling units $\rho_i$ in cell $i$, but that of the time-varying real-time density $\rho_i(t)$ in cell $i$, at any time $t$.
Estimating dynamic population densities is more challenging than approximating the static distribution of dwelling units, due to the much shorter timescale of people movements (minutes) compared to that of domicile variation (years).

The main problem in estimating dynamic populations is the lack of ground-truth data, which makes training supervised models such as that in (\ref{eq:estimreg}) impossible. Simply reusing the parameters $\hat{\alpha}$ and $\hat{\beta}$ computed for the static population is inaccurate, because those values describe the relationship between a specific subscriber presence $\sigma_i$ and the static density of residents: there is no certainty that the correspondence remains valid when inhabitants are not at home. For instance, the reduced correlation between census population information and mobile subscriber presence during working hours, in Figure~\ref{fig:hetero}, already advises against using parameters trained over night hours to estimate whole-day population densities at any time $t$.

Our approach roots instead in a novel multivariate relationship between the population density, the subscriber presence and the level of mobile communication activity of subscribers. This allows taking the constants $\alpha$ and $\beta$ out of the equation, and finding a unifying expression that can be used to estimate dynamic urban populations in absence of ground-truth information.

\vspace*{-8pt}
\subsection{Subscriber presence and activity levels}
\label{sub:dyn-activity}

\begin{figure*}[tb]
\centering
\begin{minipage}[b]{0.3\linewidth}
	\centering
	\includegraphics[width=1\linewidth]{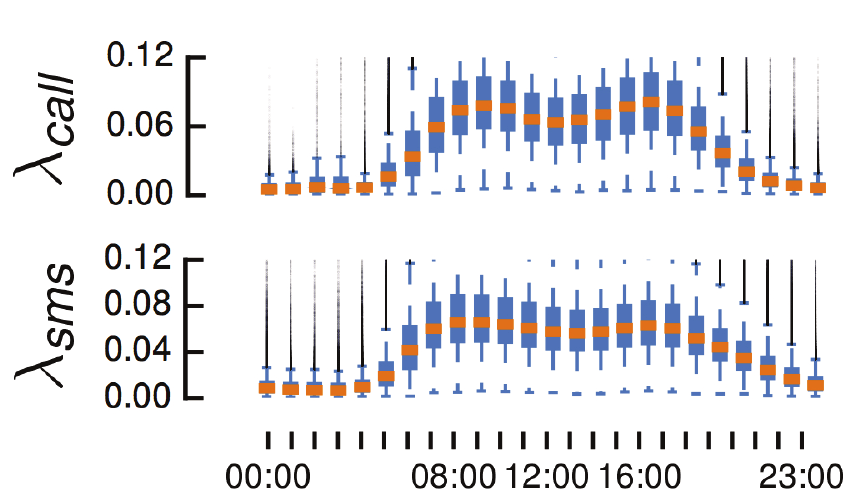}
	\small{Milan}
\end{minipage}
\hspace*{-10pt}
\begin{minipage}[b]{0.3\linewidth}
	\centering
	\includegraphics[width=1\linewidth]{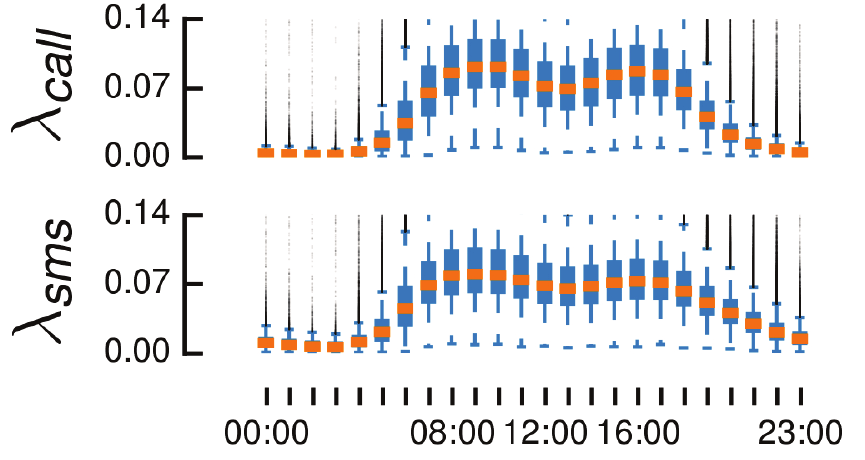}
	\small{Rome}
\end{minipage} 
\hspace*{-10pt}
\begin{minipage}[b]{0.3\linewidth}
	\centering
	\includegraphics[width=1\linewidth]{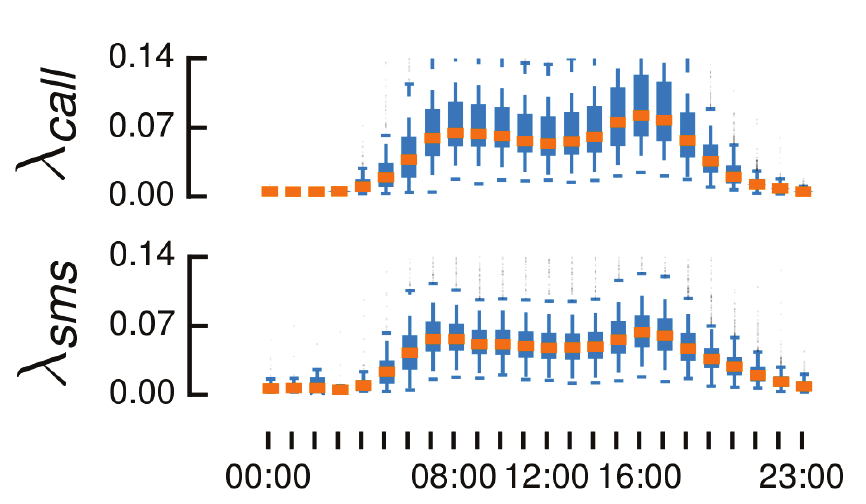}
	\small{Turin}
\end{minipage} 
\vspace*{-2pt}
\caption{Milan, Rome, Turin. Daily activity level of mobile subscribers in residential areas, for calls (top) and texts (bottom).}
\label{fig:activity}
\vspace*{-14pt}
\end{figure*}

We start by discussing the interplay between the time-varying subscriber presence $\sigma_i(t)$ and the subscriber activity level. The latter is the frequency at which a subscriber interacts with the mobile network.
Formally, we define the activity levels for voice calls $\lambda^{call}_i(t)$ and text messages $\lambda^{sms}_i(t)$ at network cell $i$ and time $t$ as:
\begin{eqnarray}
\lambda^{call}_i(t) & = & \frac{\mathcal{V}^{callin}_i(t) + \mathcal{V}^{callout}_i(t)}{\sigma_i(t)}
\label{eq:activity-call-i} \\
\lambda^{sms}_i(t) & = &  \frac{\mathcal{V}^{smsin}_i(t) + \mathcal{V}^{smsout}_i(t)}{\sigma_i(t)},
\label{eq:activity-sms-i}
\end{eqnarray}
where $\mathcal{V}^{\star}_i(t)$ stands for the number of mobile communication events of type $\star$ (\textit{i.e.}, incoming or outgoing voice calls, incoming or outgoing text messages) recorded in mobile network cell $i$ at time $t$. The subscriber presence $\sigma_i(t)$ is a proxy of the number of ``unique users'', as it provides an approximate tally of the mobile devices based on their interactions with the mobile network: the fractions in (\ref{eq:activity-call-i}) and (\ref{eq:activity-sms-i}) respectively denote the mean number of calling and texting events per user in cell $i$ at time $t$.

We can now introduce the average activity levels as 
\begin{eqnarray}
\lambda_{call}(t) & = & \frac{1}{N} \sum_{i=0}^{N} \lambda^{call}_i(t)
\label{eq:activity-call} \\
\lambda_{sms}(t) & = & \frac{1}{N} \sum_{i=0}^{N} \lambda^{sms}_i(t),
\label{eq:activity-sms}
\end{eqnarray}
where $N$ denotes the number of mobile network cells in the target geographical region.
Then, $\lambda_{call}(t)$ and $\lambda_{sms}(t)$ express the average number of calls and messages sent or received at time $t$ by a user located in the whole urban areas.
    
These activity levels are not uniform over time. Figure~\ref{fig:activity} depicts the fluctuation of $\lambda_{call}(t)$ and $\lambda_{sms}(t)$ in the residential areas of Milan, Rome and Turin, over a day. The error bars indicate the average and 5\textsuperscript{th}, 25\textsuperscript{th}, 75\textsuperscript{th} and 95\textsuperscript{th} percentiles of the activity levels, on a hourly basis. For all cities, $\lambda_{call}(t)$ and $\lambda_{sms}(t)$ undergo major variations, with minima at night and higher activity during work hours.

\begin{figure}[tb]
	\centering
  \begin{minipage}[b]{0.495\linewidth}
    \centering
    \includegraphics[width=1\linewidth]{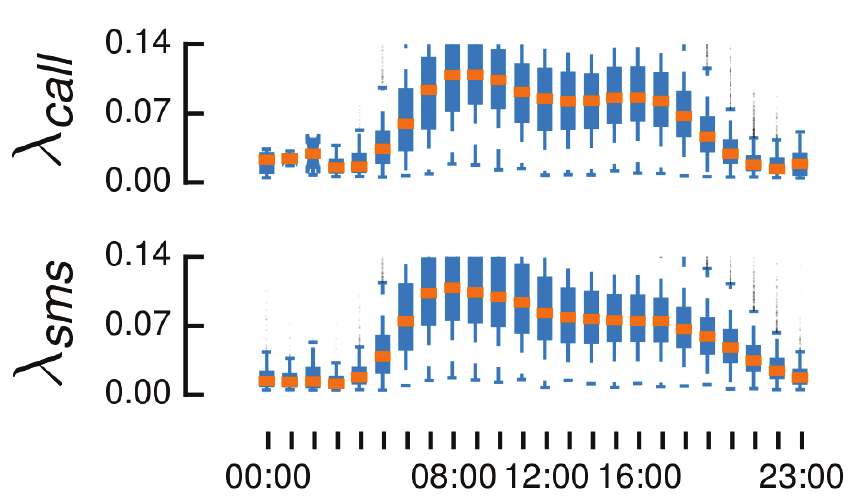} \\
      \centering{\small{Touristic}}
  \end{minipage}
	\begin{minipage}[b]{0.495\linewidth}
    \centering
    \includegraphics[width=1\linewidth]{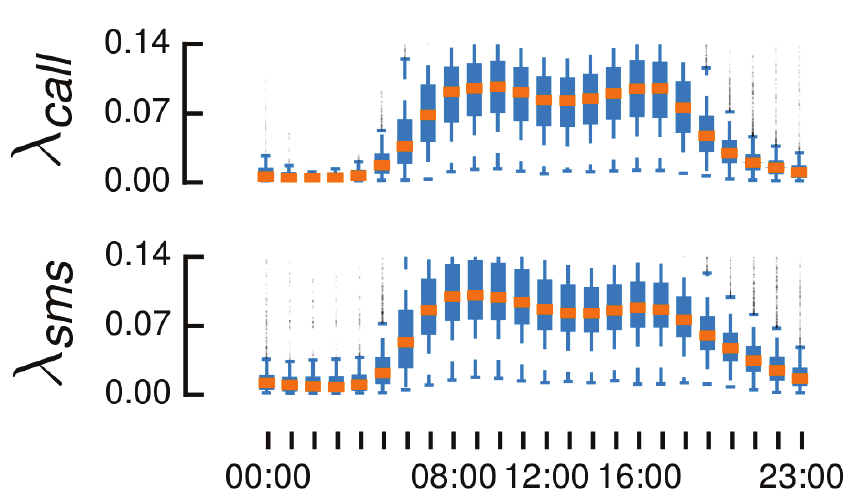} \\
      \centering{\small{University}}
  \end{minipage} 
  \\
	\begin{minipage}[b]{0.495\linewidth}
    \centering
    \includegraphics[width=1\linewidth]{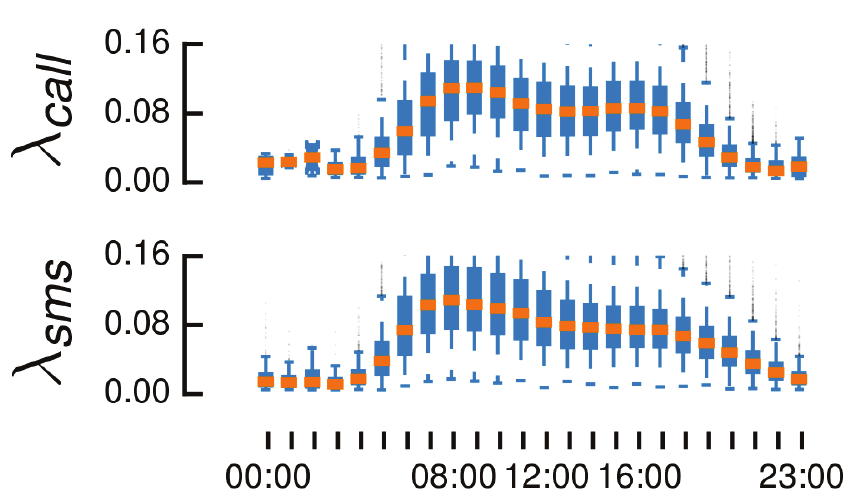} \\
      \centering{\small{Office}}
  \end{minipage} 
	\begin{minipage}[b]{0.495\linewidth}
    \centering
    \includegraphics[width=1\linewidth]{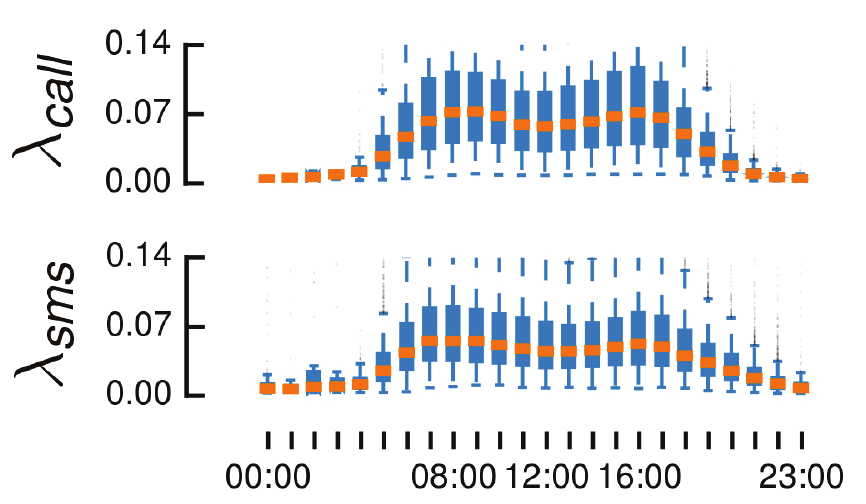} \\
      \centering{\small{Shopping}}
   \end{minipage}
	\\
	\vspace*{-5pt}
	\caption{Milan. Daily activity level of mobile subscribers in different land-use areas for calls (top) and texts (bottom).}
	\label{fig3:alllanduse}
	\vspace*{-8pt}
\end{figure}

\begin{figure}[tb]
	\centering
	\hspace*{-10pt}
  \includegraphics[width=1.05\linewidth]{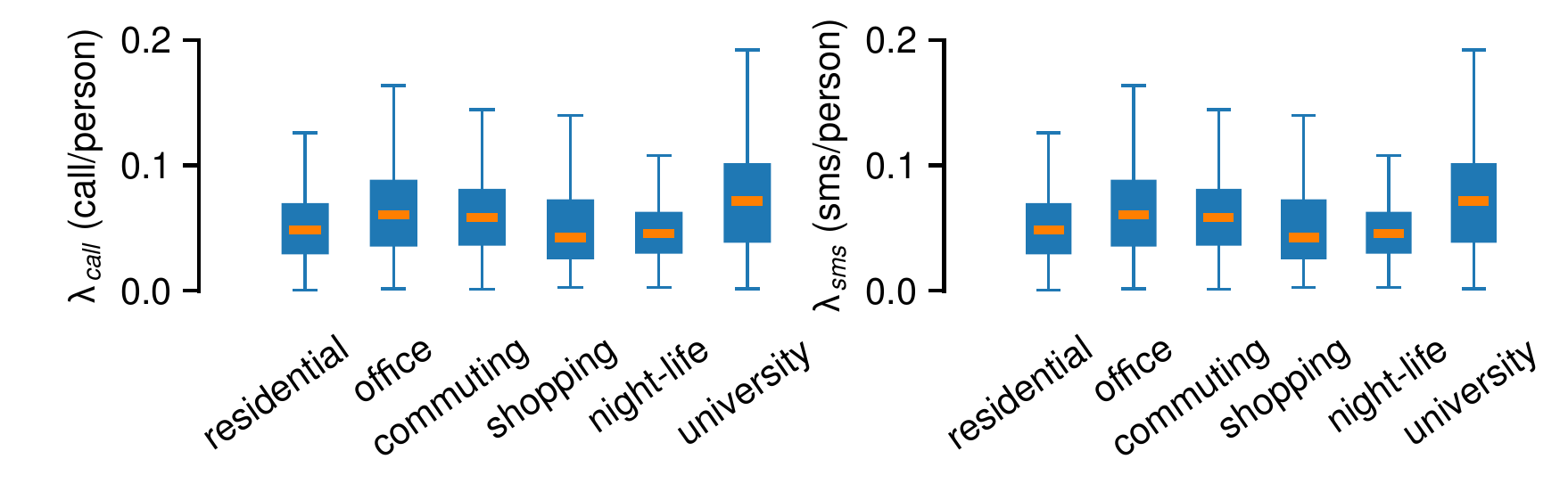}
	\vspace*{-18pt}
  \caption{Milan. Aggregate activity level of mobile subscribers in different land use areas for calls (left) and texts (right).}
  \label{fig3:activlanduse1}
	\vspace*{-8pt}
\end{figure}

This behavior is consistent across land uses, as shown in Figure~\ref{fig3:alllanduse} for the specific case of Milan. Minor variations exist, in accordance to the prevalent activities in each land land use: university and shopping areas show higher mobile communication during rush hours, and almost no network usage at night; touristic and office areas have activity peaks in the morning that then degrade until midnight. Still, the overall heterogeneity of $\lambda_{call}(t)$ and $\lambda_{sms}(t)$ is the same for all land uses: Figure~\ref{fig3:activlanduse1} highlights such uniformity, by aggregating the daily behavior into a single error bar for each land use.
In all cases, the 75-th and 95-th percentiles of both voice calls and text messages are approximately 50\% and 300\% higher than the mean activity level.
	
Our key point here is that the heterogeneity of $\lambda_{call}(t)$ and $\lambda_{sms}(t)$ in time has an impact on the correctness of the subscriber presence information. As an illustrative example, let us consider again Figure~\ref{fig:presence}: the more often a mobile device issues and receives voice calls or text messages, the more accurate is its localization in the presence data.
A legitimate question is then whether the different activity levels we observe in all our urban scenarios can be linked to the model parameterization, and explain the diversity of $\hat{\alpha}$ and $\hat{\beta}$ noted in Section~\ref{sec:STATIC_EVALUATION}.
We explore this possibility next.

\vspace*{-8pt}
\subsection{Population estimation with activity levels}
\label{sub:dyn-estimation}

We do not have access to the real values of the parameters in (\ref{eq:regression}), but to their estimations $\hat{\alpha}$ and $\hat{\beta}$.
We thus collect data in all cities that refer to the overnight period, \textit{i.e.}, from midnight to 8 am: in this period the ISTAT census information can be still considered a reliable ground truth, as most people are at home.
We then draw a scatterplot of the activity levels $\lambda_{call}(t)$ and $\lambda_{sms}(t)$, with the corresponding $\hat{\alpha}$ and $\hat{\beta}$ obtained with the RANSAC regression model.

\begin{figure*}
\begin{minipage}[t]{0.65\textwidth}
	\centering
	\hspace*{-15pt}
	\includegraphics[width=0.52\textwidth]{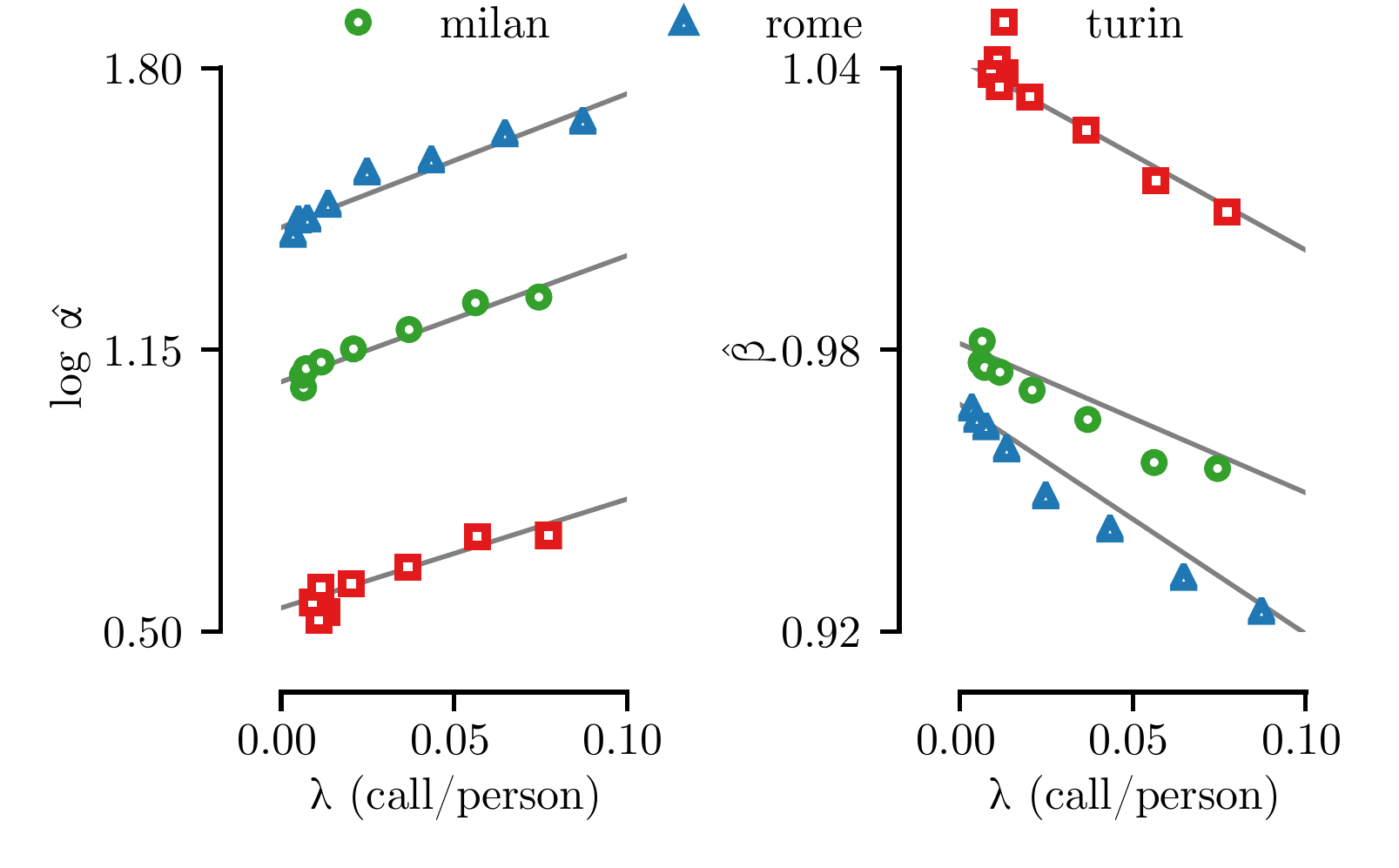}
	\hspace*{-15pt}
	\includegraphics[width=0.52\textwidth]{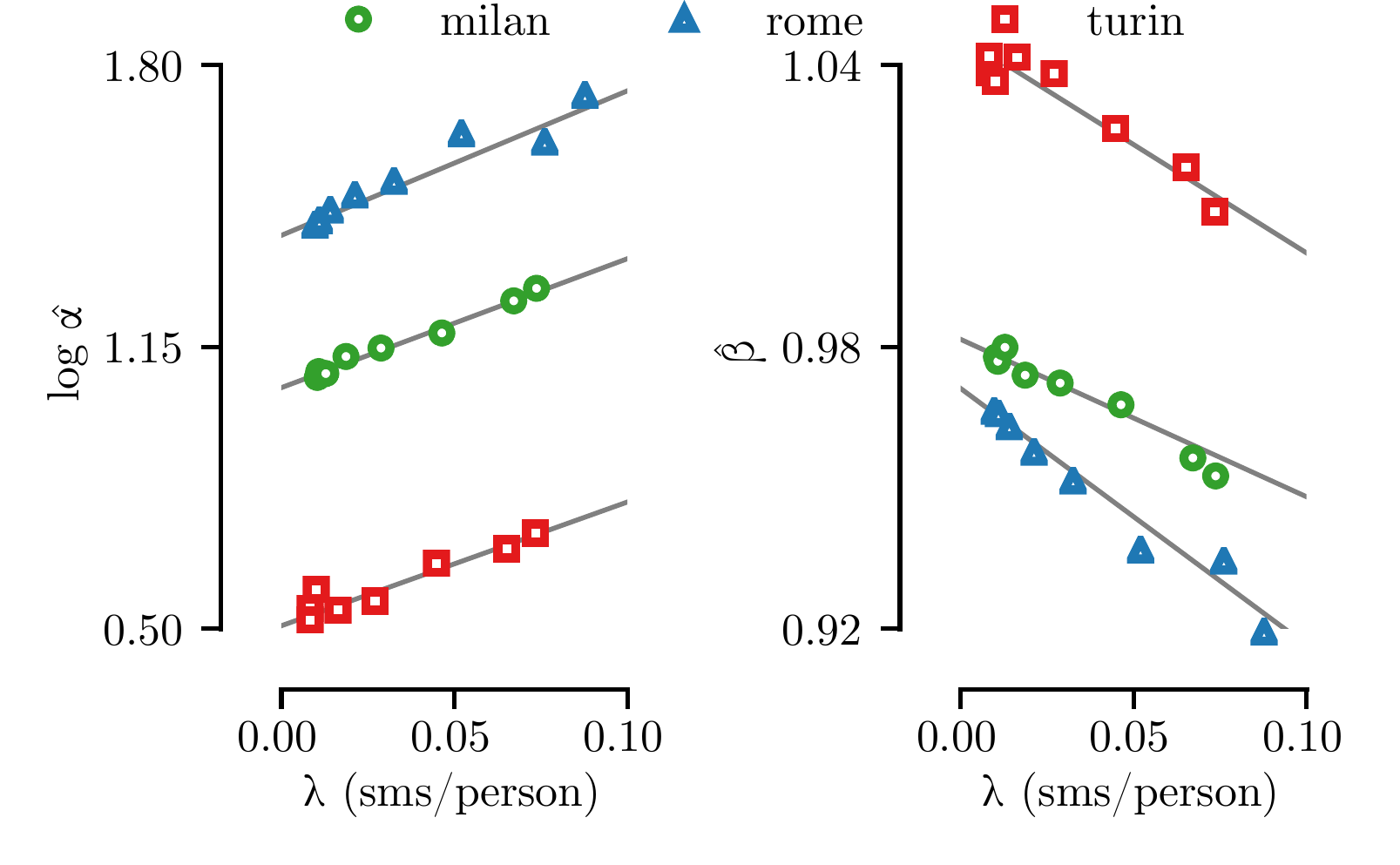}
	\vspace*{-7pt}
	\caption{Milan, Rome, Turin. Linear relationships between the activity level $\lambda_{call}$ (left)
	         and $\lambda_{sms}$ (right), and the population density model parameters $\hat{\alpha}$
	         and $\hat{\beta}$.}%
	\label{fig2:alphabeta}%
\end{minipage}
\hfill
\begin{minipage}[b]{0.32\textwidth}
	\centering
	\captionof{table}{Milan, Rome, Turin. Parameters of the multivariate model in (\ref{eq:dynamic})
	                  inferred from call and text activity.\vspace*{-3pt}}%
	\fontsize{8}{5}\selectfont%
	\renewcommand{\arraystretch}{2}
	\setlength{\tabcolsep}{1.5pt}
	\begin{tabular}{@{}cccccccrrrrrrrrr@{}}
	& \multicolumn{3}{c}{Voice calls} && \multicolumn{3}{c}{Text messages} \\
	\cmidrule{2-4} \cmidrule{6-8}
	\multicolumn{1}{c}{\color{white}------} &
	\multicolumn{1}{c}{Milan} & \multicolumn{1}{c}{Rome} & \multicolumn{1}{c}{Turin} &&
	\multicolumn{1}{c}{Milan} & \multicolumn{1}{c}{Rome} & \multicolumn{1}{c}{Turin} \\
	\midrule
	\textsc{$\hat{a}_\alpha$} & 2.90 & 3.15 & 2.34 && 2.91 & 3.11 & 2.64 \\
	\textsc{$\hat{b}_\alpha$} & 1.07 & 1.42  & 0.55 && 1.05 & 1.40 & 0.52 \\
	\textsc{$\hat{a}_\beta$} & -0.30 & -0.50 & -0.44 && -0.35 & -0.48 & -0.43 \\
	\textsc{$\hat{b}_\beta$} & 0.98 & 0.96 & 1.04 && 0.98 & 0.97 & 1.04 \\
	\bottomrule  
	\end{tabular}
	\label{tab:cities2}
	\vspace*{-23pt}
\end{minipage}
\vspace*{-8pt}
\end{figure*}

The results for the three cities are depicted in Figure~\ref{fig2:alphabeta}, and outline a striking linear relationship between the activity levels and both model parameters, for all cities.
Specifically, $\hat{\alpha}$ grows linearly with the activity levels, while $\hat{\beta}$ drops linearly with the same measures.
We can then re-write the parameters $\hat{\alpha}$ and $\hat{\beta}$ as
\begin{equation}
\hat{\alpha} = \hat{a}_\alpha \lambda_\star(t)+\hat{b}_\alpha
\label{eq:alpha-ext}
\end{equation}
\begin{equation}
\hat{\beta} = \hat{a}_\beta \lambda_\star(t) + \hat{b}_\beta,
\label{eq:beta-ext}
\end{equation}
where $\star$ denotes the type of event (\textit{i.e.}, calls or texts).

The exact values of $\hat{a}_\alpha$, $\hat{b}_\alpha$, $\hat{a}_\beta$, and $\hat{b}_\beta$ are listed in Table\,\ref{tab:cities2}.
In both (\ref{eq:alpha-ext}) and (\ref{eq:beta-ext}), we observe some heterogeneity across cities. Specifically, the derivatives of the curves are quite close to each other, whereas a constant offset tells apart the linear behavior observed in different urban areas. Such a diversity in the parameter settings evidences the need for a per-city calibration of the model.
Instead, there is no significant difference between the values obtained when considering voice call or text message activity: hereinafter, we will consider the former only.

We leverage the results above to draw a unifying multivariate model that links the population density to both the subscriber presence and the subscriber activity level. Specifically, we refine our estimation model as
\begin{equation}
\hat{\rho}_i(t) =
e^{(\hat{a}_\alpha \lambda_i(t)+\hat{b}_\alpha)} \cdot
\sigma_i(t)^{(\hat{a}_\beta \lambda_i(t) + \hat{b}_\beta)},
\label{eq:dynamic}
\end{equation}
where $\lambda_i(t)$ is a simplified notation for $\lambda^{call}_i(t)$.

The following important considerations are in order.
First, the expression in (\ref{eq:dynamic}) employs variables $\sigma_i(t)$ and $\lambda_i(t)$ that can be computed from mobile network metadata at any time $t$.
Second, unlike the original $\hat{\alpha}$ and $\hat{\beta}$ in (\ref{eq:regression}), the new parameters $\hat{a}_\alpha$, $\hat{b}_\alpha$, $\hat{a}_\beta$, and $\hat{b}_\beta$ can be regarded as time-invariant, assuming that the linear relationships in (\ref{eq:alpha-ext}) and (\ref{eq:beta-ext}) hold for any activity level.
When considered jointly, these observations imply that the model in (\ref{eq:dynamic}) is suitable for the dynamic estimation of population densities in practical cases where ground-truth information on the instantaneous distribution of inhabitants is unavailable.

\vspace*{-8pt}
\subsection{Properties of the multivariate model}
\label{sub:dyn-properties}

\begin{figure}[tb]
  \centering
	\includegraphics[width=0.85\columnwidth]{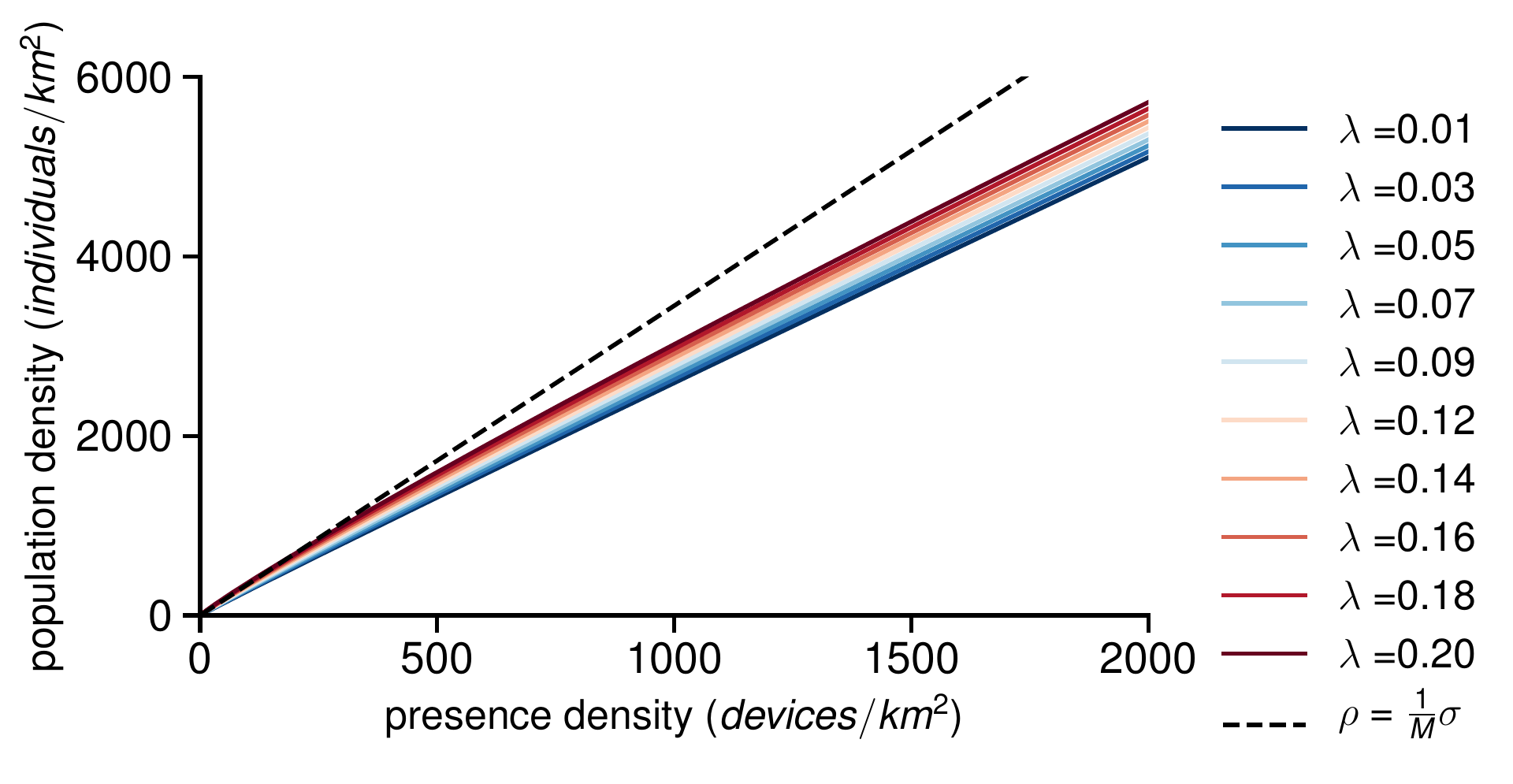}
	\vspace*{-5pt}
	\caption{Milan. Dynamic population density estimated by the multivariate model
	as a function of presence, for different values of the activity level. The dashed line
  depicts the perfectly proportional relationship $\rho_i(t)=\frac{1}{M}\cdot\sigma_i(t)$.}
	\label{fig:convergence}
	\vspace*{-8pt}
\end{figure}

We explore the basic properties of the multivariate model in Figure~\ref{fig:convergence}.
The plot illustrates how expression (\ref{eq:dynamic}) reshapes the relationship between the presence density and the estimated population density depending on the activity level $\lambda$, in the exemplary Milan scenario.
The dashed line in the figure is the conceptual curve that would link the estimated population $\hat{\rho}_i(t)$ to the presence $\sigma_i(t)$ if the latter metadata reported the exact number of customers in region $i$. In that case, the population could be obtained by simply scaling the presence by the market share $M$ of the mobile operator%
\footnote{TIM, our mobile network metadata provider, has a market share of approximately 35\% in Milan, hence $M=0.35$ in our case.},
\textit{i.e.}, $\hat{\rho}_i(t)=\frac{1}{M}\cdot\sigma_i(t)$.

We observe that the model always lies below such a conceptual curve, compensating for the effect that the presence metadata computed with current activity levels tends to overestimate the population density. However, as the value of $\lambda$ grows from 0.01 to 0.20 (the two extreme values observed in our datasets), the model approaches the ideal $\hat{\rho}_i(t)=\frac{1}{M}\cdot\sigma_i(t)$ relationship. This is consistent with the intuition that higher subscriber activity results in presence metadata that approximates more accurately the actual number of people present in a cell. Interestingly, such a correspondence occurs faster for low-density cells (\textit{e.g.}, presence is an excellent proxy for population densities below 50 individuals/km\textsuperscript{2} already at $\lambda=0.14$); instead, high-density cells require subscriber presence values that are never attained in our data. At the maximum activity level recorded in our metadata, \textit{i.e.}, $\lambda=0.20$, the model is nearly equivalent to the perfectly proportional representation for presence densities up to 400 devices/km\textsuperscript{2}.

Finally, we comment on the suitability of the model for real-time estimation of dynamic population densities. Our multivariate model has a close form, presented in (\ref{eq:dynamic}). Therefore, the computational complexity of the model corresponds to that of computing the equation -- an operation performed in nanoseconds by any low-end CPU. As a result, the complexity of the model easily meets
the requirements of any real-time application. Instead, the actual system latency would depend from the time needed by the mobile network operator to collect and process the data required to compute the
presence metadata and the activity level: however, these aspects concern the overall mobile network architecture, and are well beyond the scope of our contribution.

\vspace*{-8pt}
\section{Case studies}
\label{sub:DYNAMIC_EVALUATION}

In this section, we provide proof-of-concept exploitations of the model in (\ref{eq:dynamic}).
Specifically, we first leverage the model to provide a glance of the population dynamics during typical days in Milan and Turin. Then, we assess the model capacity to approximate the population density during special events, such as sport matches and public rallies.

\vspace*{-8pt}
\subsection{A day in the life of Milan}
\label{sub:a-day}

\begin{figure}[tb]
	\centering
	\includegraphics[width=1\linewidth]{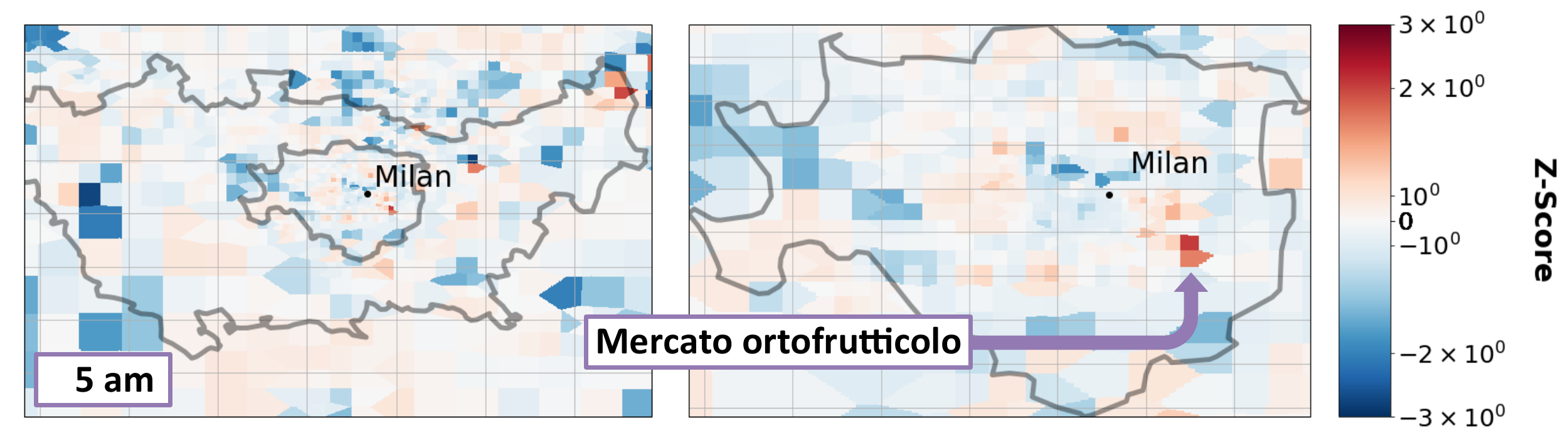}
	\includegraphics[width=1\linewidth]{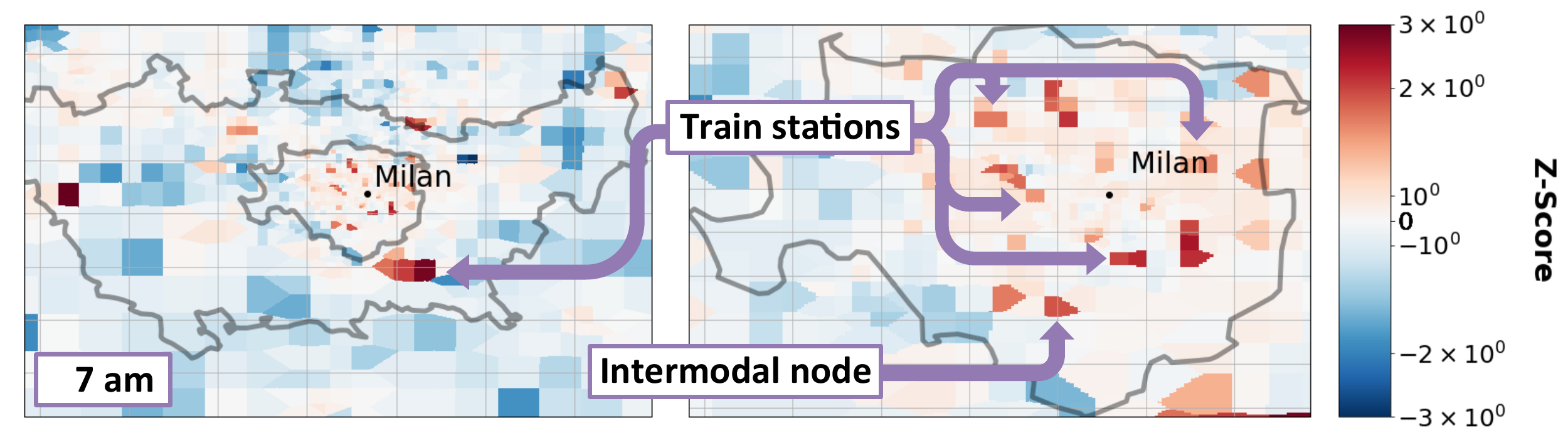}
	\includegraphics[width=1\linewidth]{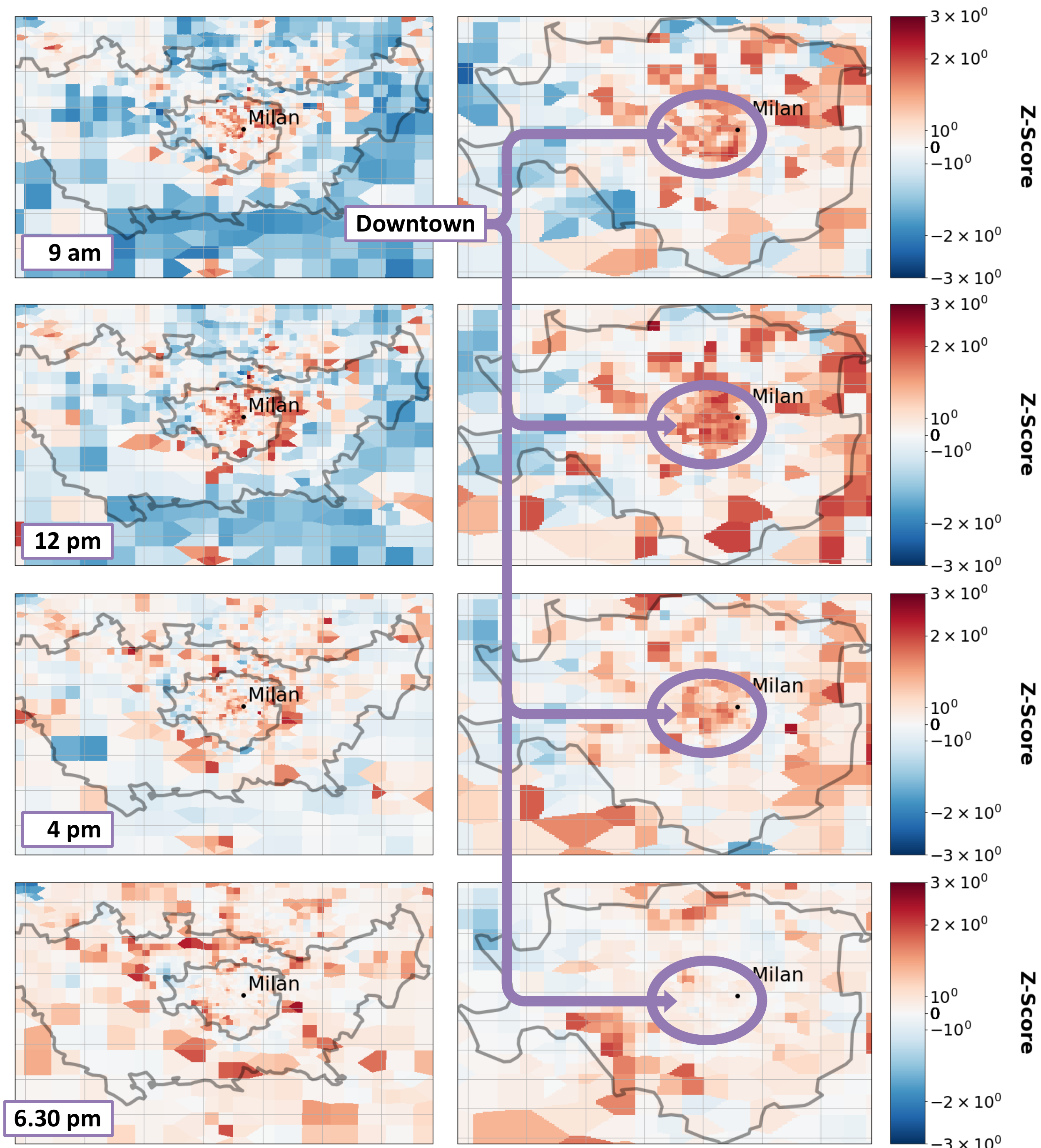}
	\includegraphics[width=1\linewidth]{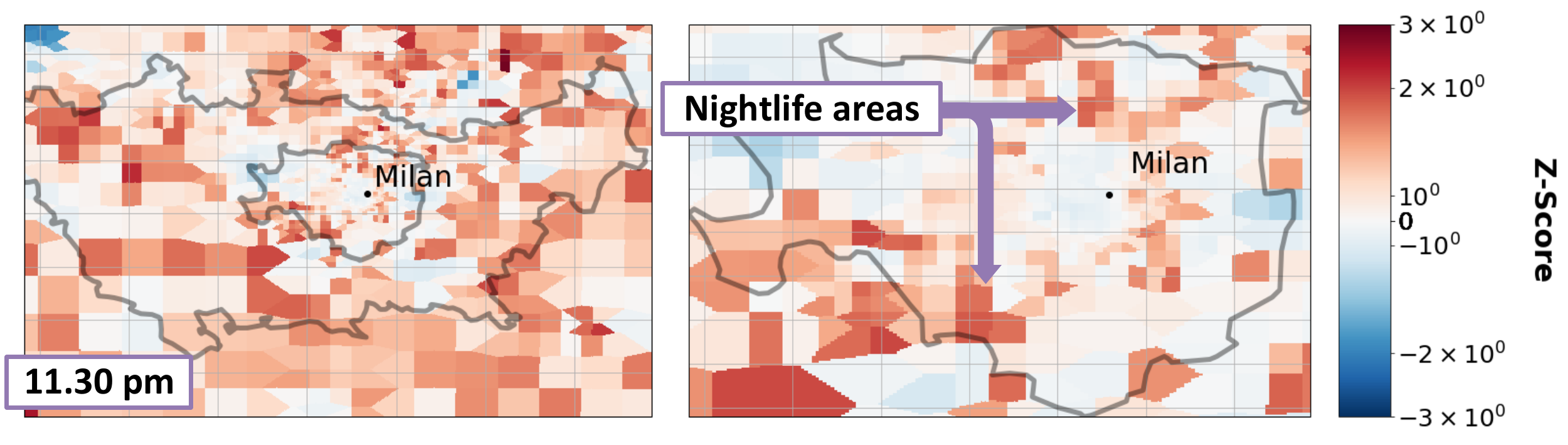}
	\vspace*{-16pt}
	\caption{Dynamic distribution of Milan population on April 13, 2015.
	Top to bottom: 5 am, 7 am, 9 am, 12 pm, 4 pm, 6.30 pm, 11.30 pm.
	Left plots show the whole conurbation, right ones the city only.
	Figure best viewed in colors.}
	\label{fig:dyn-milan}
	\vspace*{-17pt}
\end{figure}

Figure~\ref{fig:dyn-milan} illustrates the dynamic distribution of the population in Milan and its suburbs, as inferred from our multivariate model, during key times of one normal day, \textit{i.e.}, Monday April 13, 2015.
Left plots refer to the whole conurbation, whereas right plots focus on the actual city. In each plot, colors denote the z-score, which measures the deviation from the mean of the population distribution at each location. Formally, the z-score in cell $i$ at time $t$ is
\begin{equation}
z_i(t) = \frac{\hat{\rho}_i(t)-\mu_i}{\delta_i},
\label{eq:dynamic11}
\end{equation}
where $\mu_i$ and $\delta_i$ are the mean and standard deviation of the population density at cell $i$, computed over the values estimated by our model during the full two months, \textit{i.e.}, over $\hat{\rho}_i(t)$ $\forall t$.
As a result, the plots show how the population density fluctuates at specific times: variations range from high in-flow of individuals moving into a cell (red) to high out-flow of individuals leaving a cell (blue), passing by neutral cells where the population density does not vary during the observation period (white). 

Reasonable dynamics emerge from the plots. At 5 am, the only point of interest showing relevant activity is the \textit{mercato ortofrutticolo}, \textit{i.e.}, the wholesale produce market of Milan. The market attracts farmers and merchants very early in the morning, as highlighted by the red spot in the top right plot.
At 7 pm, the population density is especially high around main transportation hubs, such as train stations and intermodal exchange nodes.
The population density grows significantly in the city center throughout the morning, see \textit{e.g.}, the plots at 9 am and 12 pm. At the same times, the suburban and rural areas show low z-scores, indicating a clear in-flow of inhabitants from around the city to downtown, where office areas are located.
The trend is then reversed in the afternoon, starting at 4 pm and more clearly at 6.30 pm, when people leave from the office. Here, the city center tends to become less populated, with an out-flow of inhabitants towards the city outskirts.
Finally, it is interesting to observe that late at night, at 11.30 pm, the population is mostly at home, with high z-scores in suburban and rural areas, or in nightlife areas.

Overall, the results in Figure~\ref{fig:dyn-milan} show that our multivariate model can capture sensible dynamics of the population density at an intra-urban scale.

\vspace*{-8pt}
\subsection{Crowds at large-scale events}
\label{sub:events}

Our proposed model does not only capture typical dynamics of the urban population density, but can also estimate crowds at major social events. Interestingly, official figures about attendance at such events offer an original means to validate our assumptions on the time-invariance of the parameters in Table\,\ref{tab:cities2}, as well as the overall quality of the estimates returned by our multivariate model.

\vspace*{-8pt}
\subsubsection{Football matches}
\label{sub:event-foot}

\begin{figure}[tb]
	\centering
	\hspace*{-8pt}
	\includegraphics[width=1.05\columnwidth]{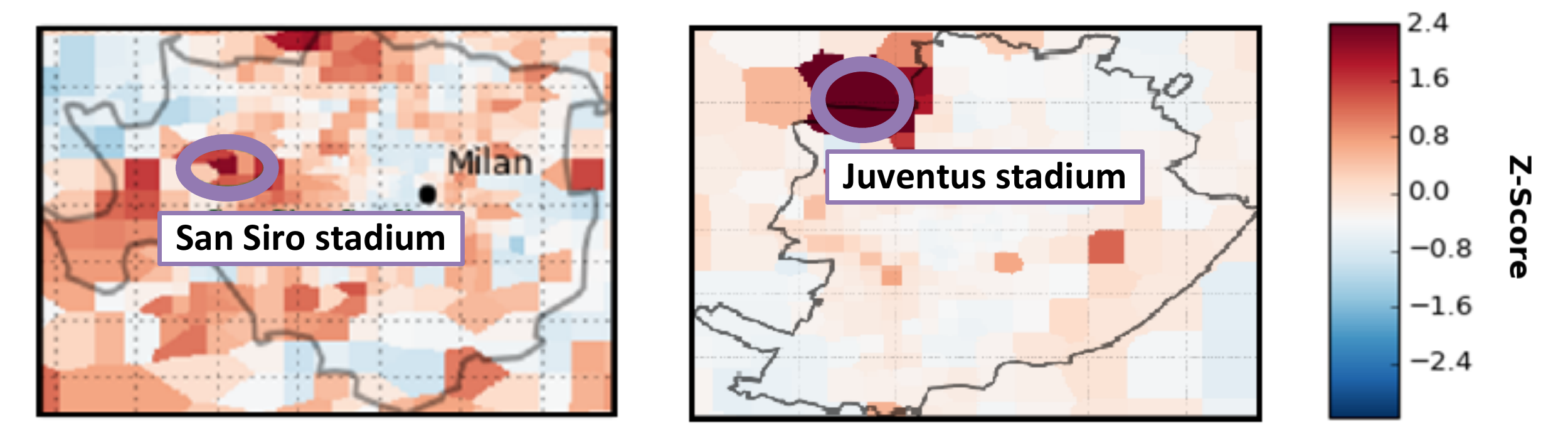}
	\vspace*{-14pt}
	\caption{Examples of dynamic distribution of populations during football matches in Milano, on April 19 at 10 pm (left), and in Turin, on April 14 at 9 pm (right).}
	\label{fig:dyn-matches}
	\vspace*{-8pt}
\end{figure}

Football matches are traditionally very popular events throughout Italy.
Figure~\ref{fig:dyn-matches} highlights the population density in Milan and Turin during games of major local football teams.
For the Milan case, the plot refers to April 19, a Sunday, at 10 pm. The model conveys well the crowd attracted by an important football match between AC Milan and Inter Milan, the two city teams, that took place on that day at San Siro, the main city stadium.
The stadium position is even more evident in the case of Turin, on April 14 at 9 pm. At that time, the local team, Juventus FC, was playing a quarter-final Champions League game against AS Monaco.

\begin{table*}[tb]
\centering 
\ra{1.1}
\caption{Comparative evaluation of estimated and actual attendance at football matches in Milan, Turin and Rome.}
\vspace*{-3pt}
\begin{tabular}{@{}cccrccccrccccr@{}}
\toprule
\multicolumn{4}{c}{\textsc{\textbf{Milan}}} &  & \multicolumn{4}{c}{\textsc{\textbf{Turin}}} &  & \multicolumn{4}{c}{\textsc{\textbf{Rome}}}\\
\cmidrule{1-4} \cmidrule{6-9} \cmidrule{11-14} 
date & $\hat{\Gamma}_s$ & visitors & error & & date & $\hat{\Gamma}_s$ & visitors & error & & date & $\hat{\Gamma}_s$ & visitors & error\\
\midrule
2015-03-01 & 36,478 & 39,310 & -7.2\% && 2015-03-05 & 36,537 & 40,211 & -9.1\% && 2015-03-02 & 54,792 & 55,651 & -1.5\% \\
2015-03-19 & 47,613 & 42,041 & 13.2\% && 2015-03-09 & 31,743 & 37,506 & -15.3\% && 2015-03-04 & 23,213 & 27,044 & -14.1\% \\
2015-03-21 & 27,658 & 30,748 & -10.0\% && 2015-03-22 & 33,390 & 39,919 & -16.3\% && 2015-03-09 & 29,682 & 33,025 & -10.1\%  \\
2015-04-04 & 26,803 & 33,175 & -19.2\% && 2015-04-04 & 31,868 & 40,029 & -20.3\% && 2015-03-16 & 29,747 & 33,250 & -10.5\% \\
2015-04-12 & 32,281 & 30,735 & 5.0\% && 2015-04-14 & 40,895 & 40,801 & 0.2\% && 2015-03-19 & 33,007 & 30,591 & 7.8\% \\
2015-04-19 & 66,761 & 74,022 & -9.8\% && 2015-04-18 & 36,113 & 38,916 & -7.2\% && 2015-04-04 & 26,248 & 34,425 & -23.7\% \\
2015-04-25 & 39,600 & 37,695 & 5.0\% && 2015-04-29 & 33,171 & 37,607 & -11.7\% && 2015-04-19 & 25,095 & 35,088 & -28.4\% \\
2015-04-29 & 26,672 & 25,916 & 2.9\% &&            &        &        & && 2015-04-29 & 27,991 & 32,367 & -13.5\% \\
\bottomrule
\end{tabular}
\label{tab:resultsmatch1}
\vspace*{-10pt}
\end{table*}

In fact, the multivariate model allows going beyond a simple visualization of population density peaks in and around the stadiums during matches. By leveraging the expression in (\ref{eq:dynamic}), we can produce actual estimates of the attendance at matches through the following steps.

\begin{itemize}

\item We identify the mobile network cells that provide coverage to the stadium, and denote their set as $\mathcal{N}$. Since exact maps of the signal propagation and antenna coverage areas are not available, we tend to be inclusive, considering all cells that intersect with the stadium surface, as well as the adjacent ones.

\item For each cell $i\in\mathcal{N}$, we determine the presence density in a normal situation, \textit{i.e.}, when no match is played at the stadium. To that end, we record all presence density values at the same weekday and time of the match, excluding those days where another match was played; we then compute the median of such presence densities, and denote the result as $\bar{\sigma}_i(t)$.

\item We establish the time of the match at which the presence density across all cells in $\mathcal{N}$ reaches its peak, \textit{i.e.},
\begin{equation}
t_{peak} = \argmax_{t\in\mathcal{T}} \sum_{i\in\mathcal{N}} \sigma_i(t),
\end{equation}
where $\mathcal{T}$ is the match timespan, from 15 minutes before kickoff to 15 minutes after the final whistle.

\item We compute the average presence density in normal conditions, $\sigma_{norm}$, and during the match, $\sigma_{match}$, as
\begin{eqnarray}
\sigma_{norm} & = & \sum_{i\in\mathcal{N}} \frac{A_i}{\sum_{j\in\mathcal{N}} A_j} \cdot \bar{\sigma}_i(t_{peak}) \\
\sigma_{match} & = & \sum_{i\in\mathcal{N}} \frac{A_i}{\sum_{j\in\mathcal{N}} A_j} \cdot \sigma_i(t_{peak}),
\end{eqnarray}
where $A_i$ indicates the surface of mobile network cell $i$. We opt for an average weighted on the relative cell sizes, so as to account for the fact that the cells in $\mathcal{N}$ may have quite diverse surfaces.

\item We calculate the mean activity level in all concerned cells during the course of match as
\begin{equation}
\widetilde{\lambda} = \frac{1}{|\mathcal{N}|} \sum_{i\in\mathcal{N}} \lambda_i(t_{peak}),
\end{equation}
where operator $|\cdot|$ designates the cardinality of a set. Here, a simple arithmetic mean suffices, since the $\lambda_i(t_{peak})$ values are averaged over users and unrelated to cell surfaces.

\item The attendance $\hat{\Gamma}$ is finally obtained via our multivariate model as
\begin{equation}
\hspace*{-5pt}\hat{\Gamma}\hspace*{-1pt}=\hspace*{-1pt}\left[(a_{\alpha} \widetilde{\lambda} + b_{\alpha}) (\sigma_{match}\hspace*{-2pt}-\hspace*{-2pt}\sigma_{norm})^{(a_{\beta} \widetilde{\lambda} + b_{\beta})} \right] \cdot\hspace*{-2pt}\sum_{j \in \mathcal{N}}A_j.
\label{eq:attendance}
\end{equation}

\end{itemize}

The last operation above applies the model in (\ref{eq:dynamic}) to the difference between $\sigma_{match}$ and $\sigma_{norm}$, \textit{i.e.}, to the increased presence density during the football match. The result is an estimate of the population density inflation (in attendees/km$^2$) caused by the crowd in the stadium. Multiplying by the total geographical surface allows inferring the actual attendance at the event.

In order to assess the quality of the estimation, we consider all matches played in March and April 2015 by the first-division football teams of Milan, Turin and Rome, and compute their attendance $\hat{\Gamma}$ from (\ref{eq:attendance}). We then compare such estimated attendance to the official figures reported by local authorities, which are very precise and represent a reliable ground-truth. The overall attendance estimation error, in terms of the metrics introduced in Section\,\ref{sub:metrics} is $R^2=0.74$ and $NRMSE(1)=0.102$: these values are aligned with those for the static population estimation, and demonstrate that our multivariate model can measure population densities in a time-varying environment with similar accuracy.

Further details are provided in Table\,\ref{tab:resultsmatch1}, which reports the date, estimated and ground-truth attendance for each match. Overall, there is a good agreement between the values, with relative errors that range from 0.2\% to 28.4\%, and an average error of 11.9\%. These results are especially encouraging when considering that our approach is general, and not specifically designed for large-scale events.

\vspace*{-8pt}
\subsubsection{Public march}

\begin{figure}[tb]
	\centering
	\hspace*{-7pt}
	\includegraphics[width=1.02\linewidth]{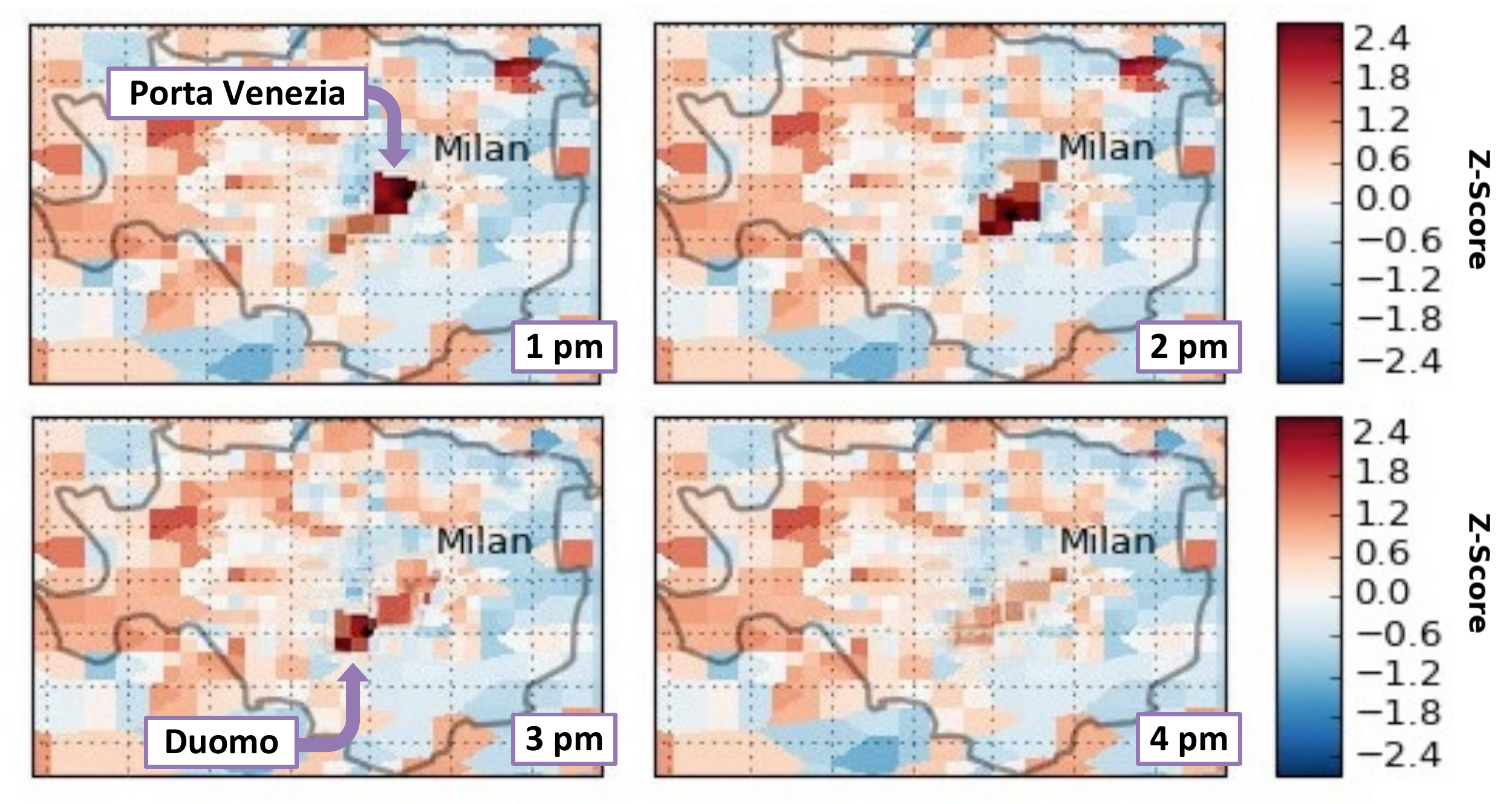} 
	\vspace*{-14pt}
	\caption{Dynamic distribution of population during a public march in Milan on April 25. Z-scores of the estimated population densities from 1 pm to 4 pm.}
	\label{fig:march}
	\vspace*{-8pt}
\end{figure}

\begin{figure}[tb]
	\centering
	\vspace*{-18pt}
	\raisebox{16pt}{
		\includegraphics[height=0.28\linewidth]{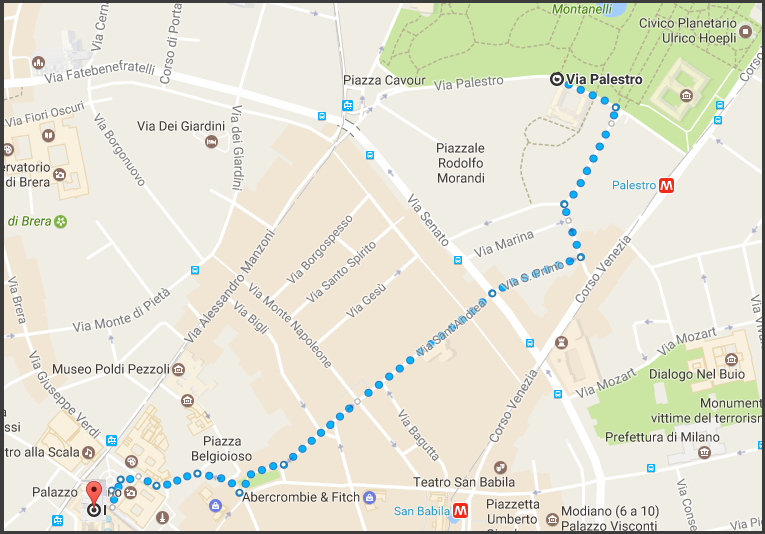}
	}
	\includegraphics[height=0.42\linewidth]{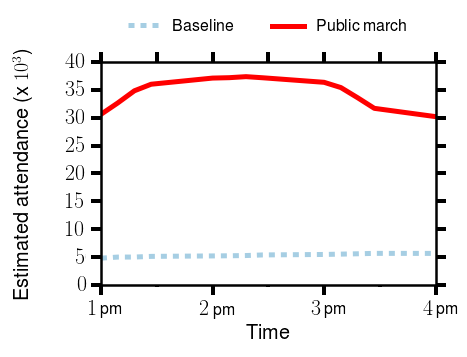}
	\vspace*{-10pt}
	\caption{Public march in Milan on April 25, 2015. Left: path of the march (blue dashed line).
	Right: estimated attendance at the march, as the usual baseline population (cyan dashed line) and increase during the march (red solid line).}
	\label{fig:mapmanif}
	\vspace*{-14pt}
\end{figure}

A second example of social happening detected from the dynamic population estimated by our multivariate model is a public manifestation that occurred in Milan, on Saturday April 25. On that day, Italy celebrated the 70th anniversary of liberation after World War II, and a large crowd marched along the streets of the city to commemorate the event. Figure~\ref{fig:march} illustrates the significant increase of the z-score of the dynamic population density inferred by our model at different times during the manifestation.
Namely, the plots allow appreciating the initial gathering of people in the Porta Venezia neighborhood, the two-hour procession in the city center, and the final arrival at the Duomo cathedral. This maps very well to the actual route of the parade, in the left plot of Figure~\ref{fig:mapmanif}.

Also in this case, our model can be leveraged to approximate the number of participants to the march. The steps are the same detailed in Section~\ref{sub:event-foot}. Notably, we consider presence metadata in all cells that provide coverage to the path of the public march in the left plot of Figure~\ref{fig:mapmanif}, as well as their immediately adjacent cells, during the whole duration of the rally.
The right plot of Figure~\ref{fig:mapmanif} shows the results returned by the multivariate model. There are approximately 5,000 people strolling in the march area during a typical early afternoon on spring Saturdays; however, the population increases dramatically on the demonstration day. We estimate the peak attendance at 37,000 persons%
\footnote{This value represents the increased presence over the normal population, as in the case of football matches.},
in the central phases of the march.
Official and unofficial figures evaluate the total number of participants at 50,000 and 60,000, respectively~\cite{25-aprile-2015-a,25-aprile-2015-b}. However, our model captures the instantaneous number of participants, and not the total one: as this is a four-hour demonstration, we speculate that the discrepancy between the official figures and our estimate is explained by a natural turnover of attendees, some of which conclude the march and leaving the manifestation before others join it at its start location.

\vspace*{-8pt}
\section{Comparative evaluation}
\label{sec:COMPARATIVE}

We carry out a comparative performance evaluation in order to clearly position
our approach with respect to previously proposed competitor solutions.
Specifically, the current state-of-the-art techniques for the estimation of
population densities from mobile network metadata are those presented
in~\cite{douglass2015high} and~\cite{xu16}. The former targets static population
distribution estimation, while the second is designed for dynamic population
density estimation. Therefore, we compare our multivariate model with the
solution in~\cite{douglass2015high} in the static case, and with that in~\cite{xu16}
in the dynamic case.

\vspace*{-8pt}
\subsection{Static population}

The approach in~\cite{douglass2015high} performs a random forest regression on
a large number of per-cell features that include the hourly volumes of incoming
and outgoing calls and text messages, the hourly volume of Internet sessions,
and the surface fraction belonging to each land use. Land uses are classified
into buildings, vegetation, water, road, and railroads, and obtained from the
OpenStreetMap database~\cite{haklay08} (41\% of the cells) or inferred from satellite
imagery (59\% of the cells). Each feature is computed at three different scales,
by considering cells in isolation as well as by aggregating neighboring cells
into 3$\times$3 and 5$\times$5 grid squares.

In fact, the random forest regression is tested in one of the scenarios
we also consider, \textit{i.e.}, the conurbation of Milan, using an equivalent
dataset provided again by TIM in the context of the 2014 Big Data Challenge.
Therefore, we can directly compare the accuracy in our results with that
attained in~\cite{douglass2015high}.
In the best configuration, where only the 16 most important features are
retained for training, the random forest technique presented in~\cite{douglass2015high}
achieves an $R^2$ of 0.66. The equivalent result obtained with our model is
0.80, as shown in Table\,\ref{tab:cities} (Milan, mixed land use): this amounts
to an improvement that exceeds 21\%.

The better performance of our model is due to a combination of factors. First,
we leverage subscriber presence, which is a more reliable proxy of population
density than the mobile network metadata used in~\cite{douglass2015high}.
Second, we employ land use information that tells apart human activities
(\textit{e.g.}, residential versus office areas) rather than simple urbanization
(\textit{e.g.}, buildings versus vegetation): therefore, our notion of land use
has a more direct relationship with population distributions.
Third, we filter the metadata based on daytime, land use and outlying human dynamics:
in this way, we account for important phenomena, such as the heterogeneity of
subscribers' behaviors over the day and during the night, the diversity of
mobile service usage in residential and non-residential area, or the variations
of mobile traffic activity during weekdays and weekends or holidays.
Given the rather intuitive nature of these phenomena, designing filters based
on reasoning is more effective than having a machine learning technique guess them.

\vspace*{-8pt}
\subsection{Dynamic population}

The approach in~\cite{xu16} mimics several of our solutions, as originally presented
in~\cite{khodabandelou}. It leverages regressions on the power-law model of population
density in (\ref{eq:regression}), where it uses the number of subscribers at 7 am as
the $\sigma_i$ variable: this is semantically similar, but not identical, to the
subscriber presence we employ as $\sigma_i$.
Also, the solution in~\cite{xu16} performs the regression on different functional
regions separately, which is equivalent to telling apart land uses. Functional
regions are urban areas characterized by different densities of points of interest,
and are classified into residence, entertainment, business, industry, education, scenery
spot and suburb.
	
However, the model in~\cite{xu16} fundamentally differs from ours in two aspects.
\begin{itemize}
\item It uses separate estimators for different functional regions. This corresponds
to regressing a different $(\alpha,\beta)$ pair for each land use in (\ref{eq:regression}).
As a result, the approach in~\cite{xu16} hinges on an array of power-law models.
\item It implements the estimation of the dynamic population density via a
time-varying rescaling factor $R_t$, which is applied uniformly to all cells%
\footnote{The study in~\cite{xu16} considers a spatial tessellation based on functional
regions and not a Voronoi tessellation based on the mobile network deployment. In our
case, we derive land use from mobile network metadata, hence the two tessellations match.}
at time $t$ and ensures that the total population stays constant over time.
\end{itemize}
Formally, the design choices above lead to a model
\begin{equation}
\hat{\rho}_i(t) = R_t \cdot \hat{\alpha}_f \: \sigma_i(t)^{\hat{\beta}_f}, \;
\text{where } \;
R_t = \frac{\sum_i \rho_i A_i}{\sum_i \hat{\rho}_i(t) A_i},
\label{eq:xu}
\end{equation}
for each spatial cell $i$ of functional region $f$ at time $t$.
In (\ref{eq:xu}), $\hat{\alpha}_f$ and $\hat{\beta}_f$ are the power-law parameters
regressed from the static population for functional region $f$, $\rho_i$ is the
ground-truth static population in cell $i$, and $A_i$ is cell $i$ surface.

\begin{figure}[tb]
	\centering
	\hspace*{-8pt}
	\raisebox{19pt}{
		\includegraphics[width=0.47\linewidth]{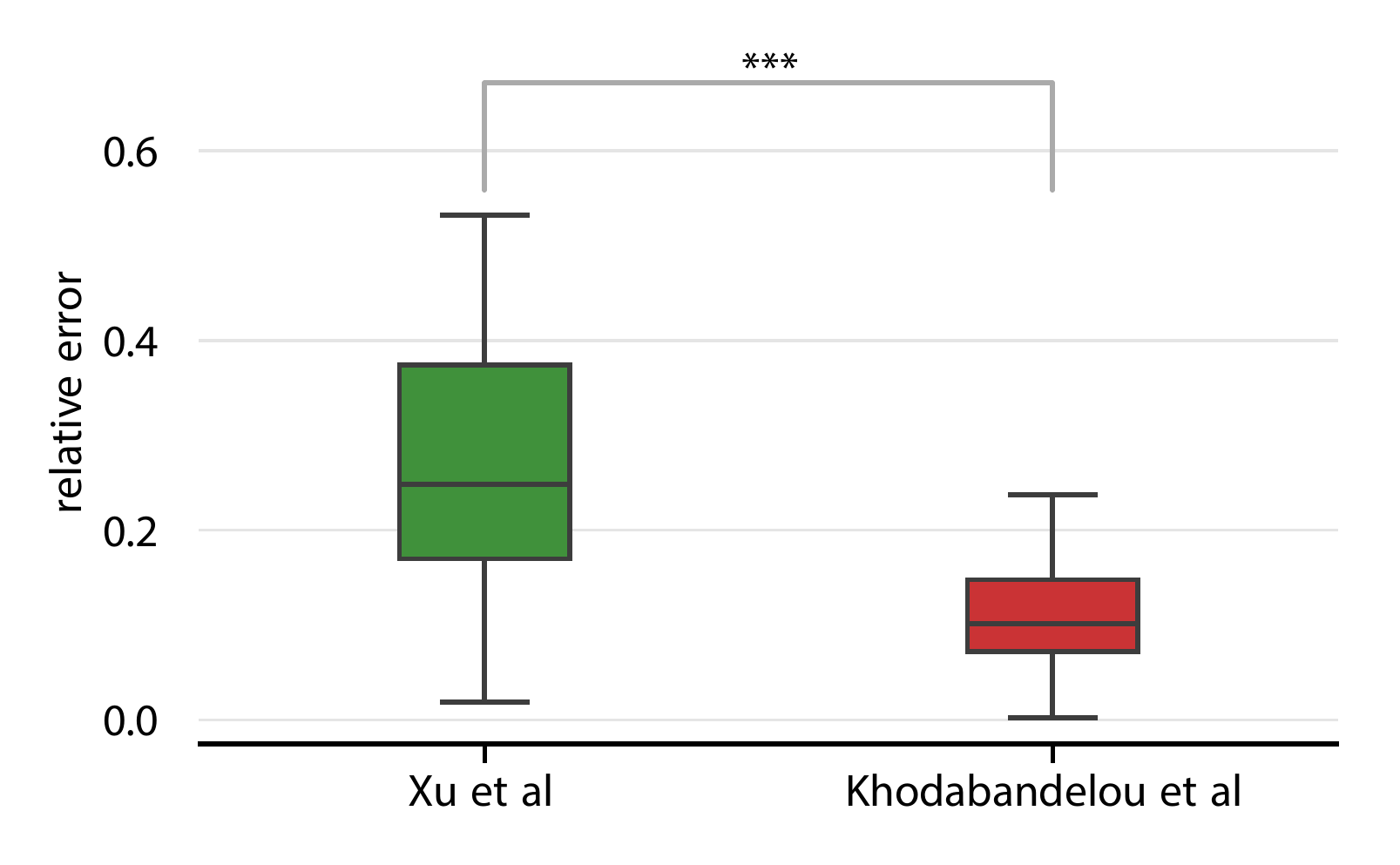}
	}
	\hspace*{-20pt}
	\includegraphics[width=0.59\linewidth]{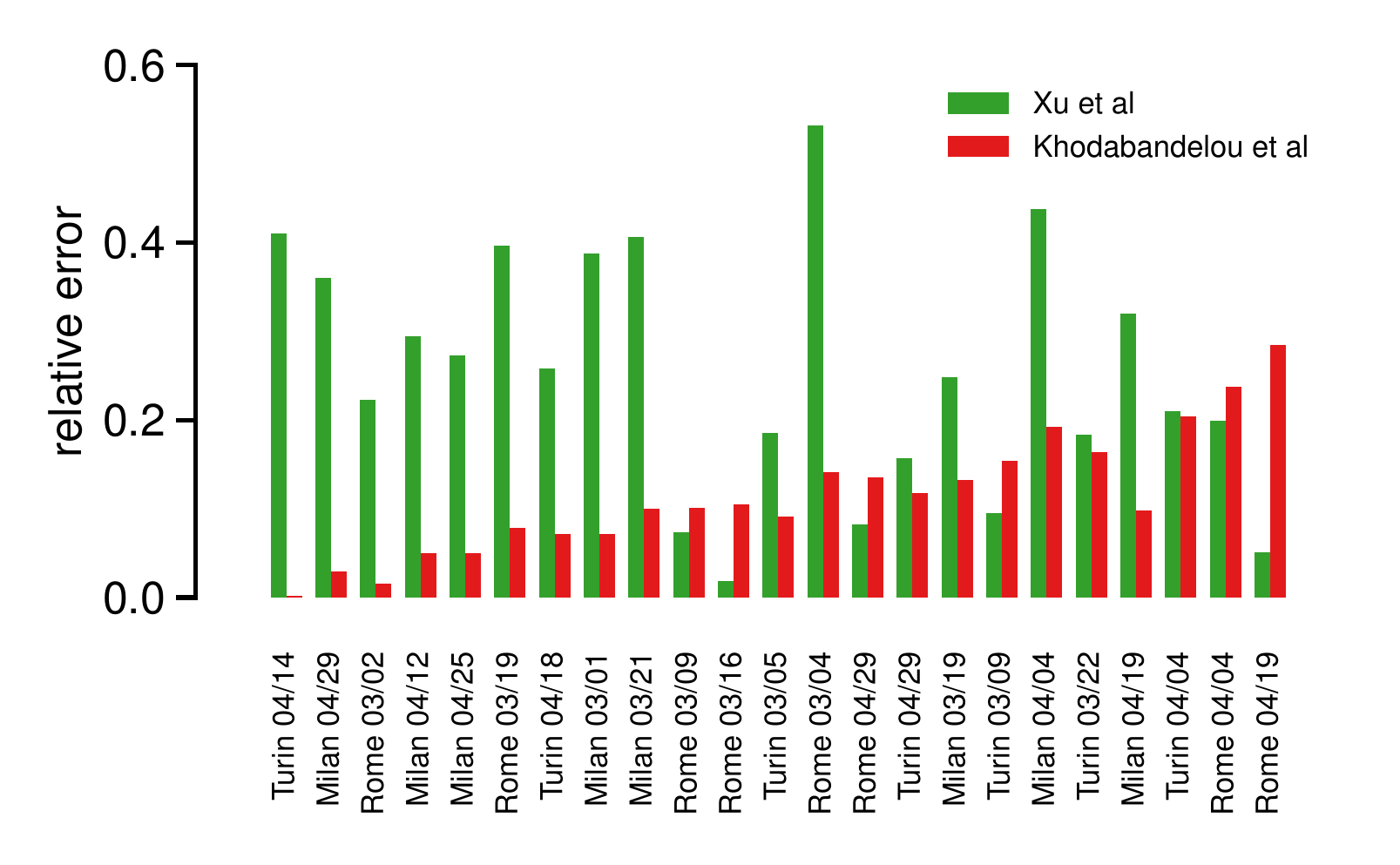}\\
	\vspace*{-4pt}
	\caption{Relative error in the estimated attendance at football matches,
	using the model in in~\cite{xu16} and our proposed multivariate model.
	Left: results aggregated over all matches in all cities.
	Right: results for individual matches.}
	\label{fig:xu_rel}
	\vspace*{-10pt}
\end{figure}

In the original work, the model in~\cite{xu16} is cross-validated with transport
data in the region of Shanghai, PRC. The urban environment and validation methodology
are very different from those we consider, hence a direct comparison with our results
is impossible.
For the sake of a fair comparison, we thus implement%
\footnote{We had to make two approximations in our implementation. First, we
use subscriber presence instead of the number of subscribers, as we do not have
access to the latter in our scenarios. Second, we employ MWS-based land uses
instead of functional regions, as we do not have access to high-detail points of
interest databases in our scenarios. We argue that these are minor changes, as
the replacement metadata is semantically close to that employed in~\cite{xu16},
and the core differences between the two models lie elsewhere.}
the solution in~\cite{xu16}, and evaluate its performance in our reference scenarios.
The validation strategy is the same we adopted in Section~\ref{sub:event-foot},
\textit{i.e.}, leveraging ground-truth attendance at football matches.

Figure~\ref{fig:xu_rel} displays the relative error of the estimation made
by the solution in~\cite{xu16} and our proposed multivariate model. The left
plot summarizes the results over all matches listed in Table\,\ref{tab:resultsmatch1},
as the 5\textsuperscript{th}, 25\textsuperscript{th}, 50\textsuperscript{th},
75\textsuperscript{th} and 95\textsuperscript{th} percentiles.
The relative error incurred by the approach in~\cite{xu16}
is approximately twice that caused by our model, for all these statistics;
in the median case, it exceeds a factor 2.2. The statistical validity of the
comparison is confirmed by a two-sided Mann-Whitney U-test, which returns
a p-value of 0.0005.
When disaggregating results for all matches, in the right plot, it is evident
that our estimation technique performs consistently better. A couple of outliers
apart, our model attains peaks of improvement at 0.4.

\begin{figure}[tb]
	\centering
	\hspace*{-8pt}
	\raisebox{19pt}{
		\includegraphics[width=0.47\linewidth]{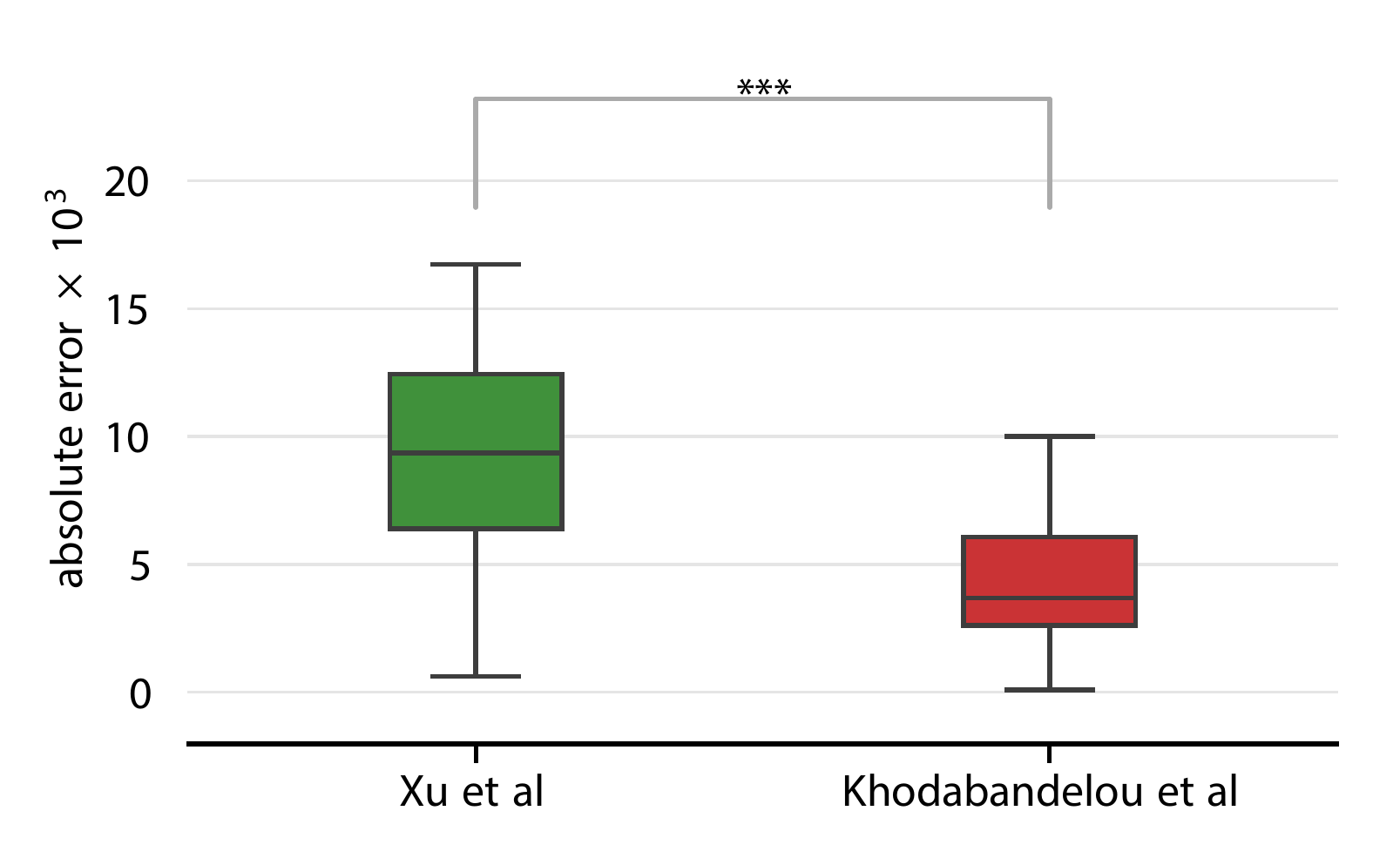}
	}
	\hspace*{-20pt}
	\includegraphics[width=0.59\linewidth]{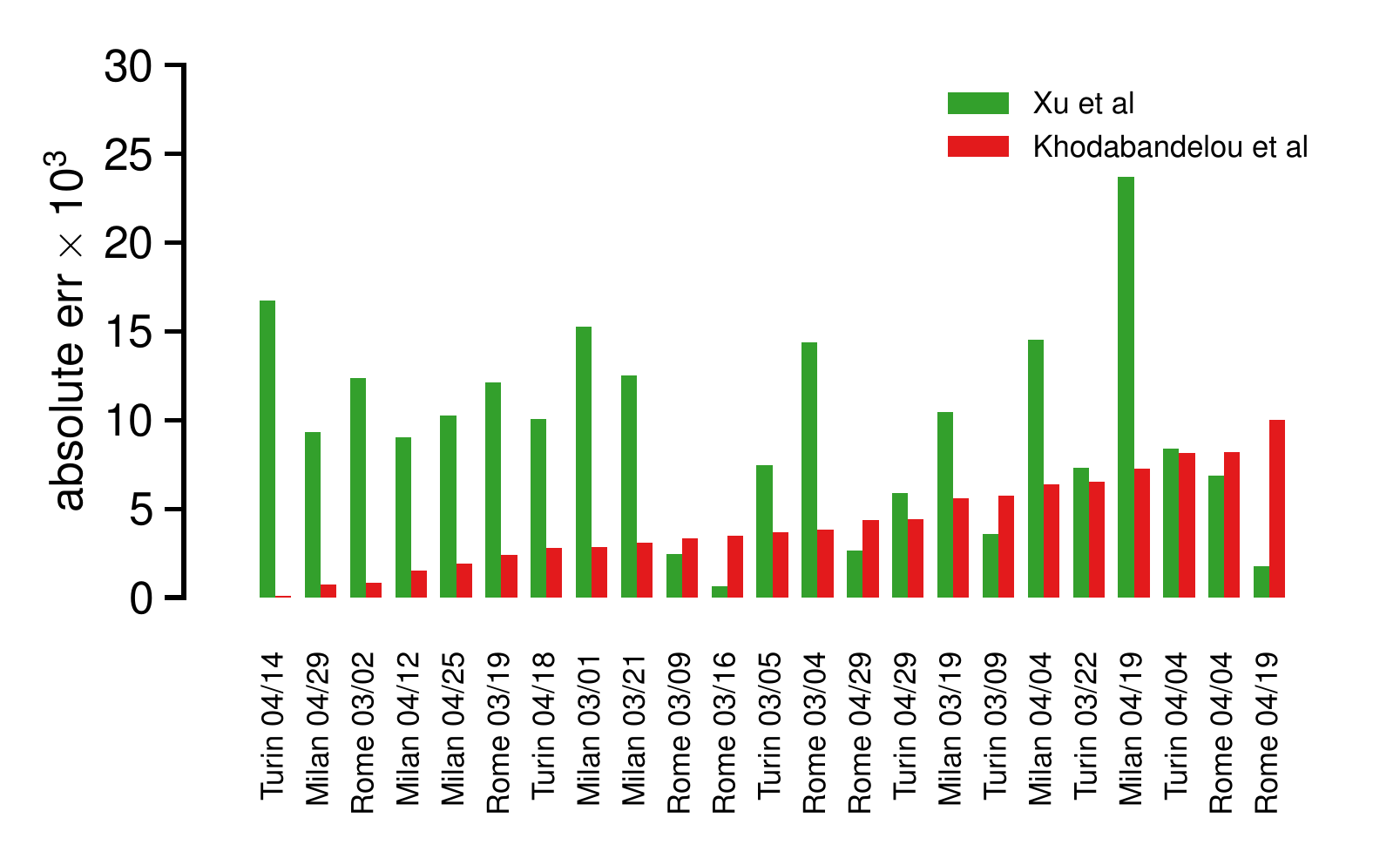}\\
	\vspace*{-4pt}
	\caption{Absolute error in the estimated attendance at football matches,
	using the model in in~\cite{xu16} and our proposed multivariate model.
	Left: results aggregated over all matches in all cities.
	Right: results for individual matches.}
	\label{fig:xu_abs}
	\vspace*{-14pt}
\end{figure}

Different matches can attract a very diverse number of spectators, hence
we also investigate how the two models compare in terms of absolute error,
\textit{i.e.}, the discrepancy in the number of attendees between estimates
and ground truth.
Figure~\ref{fig:xu_abs} shows that the performance are aligned with those
observed with relative errors: also in this case our model grants a 50\%
error reduction, consistently across matches. Thus, our conclusions
hold also in this case.

We believe that the better accuracy achieved by our model roots in
\textit{(i)} its capability to capture the dependence of the regression
parameters on the user activity levels $\lambda$, which is instead
neglected by the time-invariant $(\hat{\alpha}_f,\hat{\beta}_f)$
in~\cite{xu16}, and
\textit{(ii)} its exclusive reliance on regression in residential land uses.
Concerning this second aspect, we argue that the early morning population
in non-residential regions may not map well to people living in those areas;
depending on the nature of the area, it rather corresponds to commuters,
early workers, tourists, etc. Therefore, a regression trained on such
land uses will lead to unreliable parameters $\alpha$ and $\beta$.
As a supporting evidence from Figure~\ref{fig:xu_abs}--\ref{fig:xu_rel},
the city with more heterogeneous land uses, \textit{i.e.}, Milan, is
the one where the solution in~\cite{xu16} performs the worst.

\vspace*{-8pt}
\section{Related work}
\label{sec:RELATEDWORK}

There is wide agreement on the suitability of mobile network metadata as a source of information for human mobility analysis.
Specifically to population distribution estimation, mobile network metadata was first proposed as a proxy for the density of inhabitants in~\cite{ratti06}. Early evidences of the existence of an actual correlation between the mobile communication activity and the underlying population density were presented in~\cite{krings2009urban}, by comparing city population sizes and amounts of mobile network customers.

Subsequent works carried out more comprehensive evaluations. In~\cite{csaji2013exploring}, the home location of each subscriber was localized as the most frequently visited cell with a home profile, \textit{i.e.}, where the activity peaks at evening. The density of home locations was then found to match very well --with a 0.92 correlation-- census data on nationwide population distribution. Similarly, an excellent agreement between the overnight spatial density of mobile subscribers and that of nationwide static populations was found in~\cite{bekhor2013evaluating,calabrese2011estimating}. However, these results refer to populations at the scale of a whole country, with a spatial granularity at the level of counties or tracts, \textit{i.e.}, large regions that comprise multiple cities each.
Our focus is instead on urban population distribution estimation \textit{within} individual urban areas: down-scaling the investigation to a citywide level requires orders-of-magnitude higher accuracy, and is a harder challenge.

Citywide population estimation from mobile network metadata has been addressed by a limited number of works in the literature.
In~\cite{botta2015quantifying}, voice calls, text message, and Internet activity metadata, combined with Twitter records, were found to be highly correlated with the number of people at specific city locations (\textit{i.e.}, sports arena and airport) during a target time interval.
In~\cite{kang2012towards}, LandScan\texttrademark, a tool for ambient population estimation, was employed to explore the relationship between the voice call activity and the underlying inhabitant density, at a 1-km$^2$ resolution. The authors found a weak correlation of 0.24, later improved to 0.45 by limiting the analysis to selected time intervals rather than considering the daily communication volume.
In~\cite{deville2014dynamic}, telecommunication data was mixed with a number of other sources, including information on land use, road networks, satellite nightlights, and slope. By feeding such data to a asymmetric modeling approach, the authors obtained a high 0.92 correlation with census information. This solution is employed by the Worldpop initiative, which aims at building fine-grained maps of static population densities in underdeveloped countries~\cite{worldpop}. However, the high correlation mentioned before is a nationwide average, and the authors indicate that the accuracy is lower for the most densely populated areas, \textit{i.e.}, large cities. Indeed, in such areas, a normalized error of around 0.6 is measured in~\cite{deville2014dynamic}, whereas we obtain values below 0.1.

The current state-of-the-art in the estimation of citywide population distribution from mobile network metadata is represented by the approaches in~\cite{douglass2015high} and in~\cite{xu16}. Section~\ref{sec:COMPARATIVE} provides a thorough comparative evaluation. In addition to granting improved performance, our solution is the first that is \textit{(i)} evaluated in multiple cities so as to demonstrate its general viability, and \textit{(ii)} validated using attendances at sports and public events as ground-truth dynamic populations.

Finally, an earlier version of this paper focuses on static population estimation, and only
briefly sketches the multivariate model for the dynamic population estimation~\cite{khodabandelou}.
In addition to general presentation improvements, the present document improves the conference
version by including: \textit{(i)} a much sounder discussion of the multivariate model design;
\textit{(ii)} an original validation methodology for dynamic population estimates;
\textit{(iii)} a comprehensive evaluation of the model performance with respect to ground truth
information and in comparison with state-of-the-art benchmarks.

\vspace*{-8pt}
\section{Conclusions}
\label{sec:CONCLUSION}

We introduced a novel approach to the estimation of population density based on mobile network metadata, for both static and dynamic populations.
	
Building on the well-known power relationship between the mobile network activity density and the population density, our baseline model of static population density yields several unique properties:
\textit{(i)} it builds exclusively on metadata collected by mobile network operators avoiding complex and cumbersome data mixing;
\textit{(ii)} it leverages for the first time subscriber presence data inferred from the mobile communications of each user, which is found to be a much better proxy of the population distribution than previously adopted metrics;
\textit{(iii)} it introduces a number of original data filters based on time and land use that allow refining the population distribution estimates.
Thanks to these features, when tested with substantial real-world metadata, our model outperforms previous proposals in the literature, allowing for a reliable representation of static populations across different cities.

In addition, we extended the baseline model to a multivariate version that exploits a new time-invariant linear relationship between the power law parameters and the subscriber activity levels. The resulting multivariate model can be used to determine population densities in a dynamic fashion, as proven by evaluations with real-world mobile network metadata. Specifically, our multivariate model can reproduce typical daily activity trends in urban areas, and shows good accuracy in estimating attendance at large-scale sports and social events.

\vspace*{-8pt}
\section*{Acknowledgements}

The authors would like to thank Angelo Furno for his help with extracting land use information.
This research was partly supported by the European Union Horizon 2020 Framework Programme under REA
grant agreement no.778305 DAWN4IoE, and by BPI-France through the FUI FluidTracks project.

\bibliographystyle{IEEEtran}  
\bibliography{TMC-2017-11-0717-biblio}

\end{document}